\documentclass[12pt]{article}
\usepackage{eurosym}
\usepackage{amsfonts}
\usepackage{amssymb}
\usepackage{graphicx}
\usepackage{amsmath}
\usepackage{makeidx}
\usepackage{indentfirst}
\usepackage[T1]{fontenc}
\usepackage[utf8]{inputenc}

\newcounter{resultnum}[section]
\newcounter{conclusionnum}[section]
\newcounter{conditionnum}[section]
\newcounter{conjecturenum}[section]
\newcounter{examplenum}[section]
\newcounter{exercisenum}[section]
\newcounter{lemmanum}[section]
\newcounter{notationnum}[section]
\newcounter{theoremnum}[section]
\newcounter{definitionnum}[section]
\newcounter{corollarynum}[section]
\newcounter{remarknum}[section]
\newcounter{propositionnum}[section]
\newcounter{acknowledgementnum}[section]
\newcounter{algorithmnum}[section]
\newcounter{axiomnum}[section]
\newcounter{casenum}[section]
\newcounter{claimnum}[section]
\newcounter{summarynum}[section]
\newcounter{problemnum}[section]
\begin{document}

\title{Off-diagonal cosmological solutions in emergent gravity theories and Grigory Perelman entropy for geometric flows}
\date{December 23, 2020}
\author{ \vspace{.1 in} \textbf{Sergiu I. Vacaru} \thanks{
emails: sergiu.vacaru@gmail.com and sergiuvacaru@mail.fresnostate.edu ;%
\newline
\textit{Address for post correspondence in 2019-2020 as a visitor senior
researcher at YF CNU Ukraine is:\ }  Gagarina street, 37, ap. 3, Chernivtsi,
Ukraine, 58008} \\
%EndAName
{\small \textit{Physics Department, California State University at Fresno,
Fresno, CA 93740, USA; and }}\\
{\small \textit{Dep. Theoretical Physics and Computer Modelling, 101
Storozhynetska street, Chernivtsi, 58029, Ukraine; }} \vspace{.1 in} \\
{\ \textbf{El\c{s}en Veli Veliev} \vspace{.1 in} }\thanks{%
email: elsen@kocaeli.edu.tr and elsenveli@hotmail.com } \\
{\small \textit{Department of Physics,\ Kocaeli University, 41380, Izmit,
Turkey}} \vspace{.1 in} \\
and \vspace{.1 in} \\
\vspace{.1 in} {\ \textbf{Lauren\c{t}iu Bubuianu}}\thanks{%
email: laurentiu.bubuianu@tvr.ro } \\
{\small \textit{SRTV \ - Studioul TVR Ia\c{s}i, \ 28 Alexandru Lapu\c{s}%
neanu street, Ia\c{s}i, \ 700057, Romania;}} \\
{\small and } {\small \textit{University Apollonia, 2 Muzicii street, Ia\c{s}%
i, \ 700399, Romania }} }
\maketitle

\begin{abstract}
We develop an approach to the theory of relativistic geometric flows and emergent gravity defined by entropy functionals and related statistical thermodynamics models. Nonholonomic deformations of G. Perelman's functionals and related entropic values are used for deriving relativistic geometric evolution flow equations. For self-similar configurations, such equations describe generalized Ricci solitons defining modified Einstein equations. We analyse possible connections between relativistic models of nonholonomic Ricci flows and emergent modified gravity theories. We prove that corresponding systems of nonlinear partial differential equations, PDEs, for entropic flows and modified gravity posses certain general decoupling and integration properties. There are constructed new classes of exact and parametric solutions for nonstationary configurations and locally anisotropic cosmological metrics in modified gravity theories and general relativity. Such solutions describe scenarios of nonlinear geometric evolution and gravitational and matter field dynamics with pattern-forming and quasiperiodic structure and various space quasicrystal and deformed spacetime crystal models. We analyse new classes of generic off-diagonal solutions for entropic gravity theories and show how such solutions can be used for explaining structure formation in modern cosmology. Finally, we speculate why the approaches with Perelman--Lyapunov type functionals are more general or complementary to the constructions elaborated using the concept of Bekenstein--Hawking entropy.

\vskip1pt


\vskip3pt

PACS2010:\ 02.40.-k, 02.90.+p, 04.20Cv, 04.20.Jb, 04.50.-h, 04.90.+e,
05.90.+m

\end{abstract}

\tableofcontents

%\newpage

%\vskip10pt

\section{Introduction}

The most inspiring ideas in recent development of the gravity theory and
cosmology are on the emergent thermodynamic nature of the spacetime
geometry, when the Einstein equations can be derived using area-entropy
formulas for horizons of black holes, BH, \cite%
{bekenstein72,bekenstein73,bardeen73} and from supposed elastic properties
of gravity \cite{padmanabhan09,verlinde10,verlinde16}. A substantial
progress includes the research on the microscopic origin of
Bekenstein--Hawking entropy in string theory \cite{strominger96}, the
holographic principle \cite{susskind94,thooft93}, BH complementarity \cite%
{susskind93}, and gauge/gravity correspondence \cite{witten98,aharony99}.
Here we note a subsequent development of the (anti) de Sitter, (A)dS, and
conformal field theories, CFTs, and AdS/CFT, correspondence \cite%
{maldacena98} and formulating laws of the thermodynamics of 'apparent'
horizons \cite{ashtekar04}.

Later, it was proposed that gravity theories and generalized/ modified/
linearized Einstein equations are consequences of the quantum entanglement
connecting nearby spacetime regions \cite%
{raamsdonk10,maldacena13,lashkari13,ryu06a,ryu06}. Recent theoretical
activities are devoted to proofs that the entanglement first law and dual
gravity are derived from the CFT and/or reveal a deep connection to ideas on
emergence of spacetime and gravity from general quantum information
principles \cite%
{lloyd12,faulkner14,swingle12,jacobson15,pastawski15,verlinde16}. It was
pointed out that in a model of dual gravity with entanglement is equivalent
to the full (nonlinear) equations \cite{oh18}. An intriguing conjecture that
gravity links to an entropic force as a spacetime elasticity was proposed by
E. Verlinde \cite{verlinde10,verlinde16}. It was based on the idea that
gravitational interactions result from information regarding the positions
of material bodies. Emergent phenomena for gravity were investigated by many
other authors and their efforts involve the holographic principle in
particle physics and the information theory. There were studied geometric
models and possible applications related to gravity and quantum computers;
quantum gravity; cosmological inflation and acceleration; and dark energy
and dark matter physics etc., see \cite%
{witten18,preskill,casini11,lewkowycz13,solodukhin11} and references therein.

In the quest to explore the connection between the models of emergent
gravity and in modified gravity theories, MGTs, or general relativity, GR,
one involve a strict area law for the BH, (A)dS, or entanglement entropy and
further developments for holographic models. To derive gravitational field
equations was considered that a small but nonzero volume law entropy would
compete with, and at large distances involves, the area law. In certain
models this is due to thermalization, elastic spacetime properties, quantum
entanglement, holographic effects etc. It was proposed that such a
phenomenon occurs in the dS space being responsible for the presence of a
cosmological horizon. Nevertheless, in this series of two works, see \cite%
{partner1} as a partner paper with complementary results, we deal with quite
different issues on geometric flow modifications of gravity and spacetime
thermodynamics. It is just shown that relativistic generalizations with a
corresponding choice of evolution and thermodynamic functionals support the
ideas on the origin of gravity as an effect of the entropic force but with a
new type of geometric thermodynamics entropy. Such an approach with
Perelman--Lyapunov entropy type functionals was developed in our works on
entropic nonholonomic geometric flow evolution, nonlinear dynamics and
thermodynamics for relativistic, noncommutative, fractional, supersymmetric,
Finsler-Lagrange-Hamilton etc. generalizations of the theory of Ricci flow
evolution and applications in modern gravity and cosmology \cite%
{vacaru06,vacaru07,vacaru09,vacaru10,vacaru13,ruchin13,gheorghiu16,
rajpoot17,bubuianu18a}.

The goal of this paper is to elaborate on geometric and physical theories
relating relativistic generalizations of the Poincar\'{e}--Thurston
conjecture\footnote{%
originally formulated and proved, respectively, due to R. Hamilton and G.
Perelman, for Ricci flows of Riemannian metrics; in certain sense, this
states that our universe has the topology of a three dimensional sphere,
which is considered as one of the fundamental results in modern mathematics}%
, emergent and modified gravity constructions, and E. Verlinde conjecture
that gravity results from an entropic force as a spacetime elasticity which
explain fundamental properties of dark matter, DM, and dark energy, DE, in
modern cosmology \cite{verlinde10,verlinde16}. On topology and geometry of
Ricci flows, we refer to classical works \cite{hamilt1,hamilt2,hamilt3} and
\cite{perelman1} and reviews of mathematical results in monographs \cite%
{monogrrf1,monogrrf2,monogrrf3}). Here we note that D. Friedan published a
series of works on nonlinear sigma models, $\sigma $--models, in 2+$%
\varepsilon $ dimensions, see \cite{friedan1,friedan2,friedan3} where
geometric flow equations were introduced for the renorm group, RG, theories,
see recent results in \cite{verlinde00,tsey,verlinde13}.

The other goal of this article is to develop the anholonomic frame
deformation method, AFDM, (on early works see \cite{v05,v09,v09a} and
references therein), for constructing exact and parametric quasiperiodic
solutions of geometric entropic flow and modified gravity equations. For
reviews of recent results on black hole solutions in MGTs \cite%
{gheorghiu14,bubuianu17}, space and time like (quasi) crystals, pattern
forming and nonlinear gravitational wave structures and applications in
modern cosmology see \cite{v16,amaral17,aschheim18,bubuianu18,vacaru18tc}
and references therein. We elaborate on new classes of generic off-diagonal
stationary and cosmological solutions with entropic geometric flows which
for self similar Ricci soliton configurations result in equations considered
in E. Verlinde works \cite{verlinde10,verlinde16} and a covariant
generalization due to S. Hossenfelder \cite{hossenfelder17}, see also
critics Refs. \cite{dai17a,dai17b}.

Three lines of evidence motivate this article and the partner letter \cite%
{partner1}. First, we use our former results and nonholonomic geometric
methods \cite%
{vacaru06,vacaru07,vacaru09,vacaru10,vacaru13,ruchin13,gheorghiu16,rajpoot17,bubuianu18a}
that generalized/modified relativistic flow equations and Einstein equations
in GR and MGTs can be derived as systems of PDEs for modified Ricci solitons
for respective nonholonomic modifications of G. Perelman's F- and W-entropy
functionals. On modified gravity and applications in modern cosmology and
astrophysics, see reviews \cite{nojiri17,capozziello,hossain15,bubuianu18}.
Second, a number of recent works invoke ideas on origin of gravity as an
emergent effect of the entropic force, entanglement etc., see \cite%
{verlinde10,lloyd12,faulkner14,swingle12,jacobson15,pastawski15,verlinde16,
raamsdonk10,maldacena13,lashkari13,ryu06a,ryu06,oh18}. We argue that this
can be grounded and explained, at a deeper level, through modifications of
the Poincar\'{e}--Thurston conjecture on geometric flows, when the F- and
W-functionals are generalized for metrics and generalized connections on
Lorentz manifolds and/or certain supersymmetric/ noncommutative /
fractional/ stochastic generalizations. In our approach, the spacetime
evolution and gravity are treated via geometric entropy values which allows
to formulate respective statistical thermodynamics models. Third, new
advanced methods for constructing exact solutions in MGTs and GR allows us
to proceed directly toward definition of gravitational entropy and
thermodynamic values making no use of holography, area-entropy relation, CFT
duality etc. Due to the competition between area and volume law of
generalized W-entropy, we can characterize thermodynamically new classes of
BH and cosmological solutions with quasi-periodic structure, locally
anisotropic inflation and accelerating scenarios, exhibiting memory effects
in the form of an entropy displacement caused by matter etc.

This article is organized as follows: In Sec. \ref{s2}, we provide an
introduction to the geometry of double, 2+2 and 3+1 dimensional, spacetime
fibrations defining elastic and quasiperiodic configurations both for
gravitational and (effective) matter fields. Main concepts and most
important results on nonlinear connection geometry on nonholonomic Lorentz
manifolds, hypersurface geometric objects, and modified emergent/ elastic
gravity theories are outlined. There are studied the geometry of
distributions, and respective Lagrange densities and geometric evolution or
dynamical fields defining elastic and quasiperiodic structures.

Sec. \ref{s3} is devoted to the theory of geometric flows and modified
entropic gravity. We postulate canonical nonholonomic deformations of
Perelman's F- and W--functionals encoding geometric flow evolution scenarios
of entropic spacetimes with quasiperiodic structure. Such values are defined
in relativistic 4-d form and for 3-d hypersuface projections. There are
derived respective (generalized R. Hamilton) geometric flow equations for
entropic quasiperiodic flows. The concept of nonholonomic Ricci solitons as
self-similar configurations is elaborated and related modifications of the
Einstein equations are analyzed. We speculate on connection between
relativistic generalizations of the Poincar\'{e}--Thurston conjecture for
Ricci flows and geometric proofs of E. Verlinde's conjecture.

In Sec. \ref{s4}, we develop and apply the anholonomic frame deformation
method, AFDM, \cite%
{partner1,bubuianu18,ruchin13,gheorghiu14,rajpoot17,vacaru18tc}) in order to
prove general decoupling properties and integrability of nonholonomic
geometric flow and Ricci soliton equations encoding elastic and
quasiperiodic spacetime and (effective) matter fields properties. Such
solutions are described by generic off-diagonal metrics, and generalized
connections, depending on all spacetime coordinates and temperature like
parameters via general classes of generating functions and (effective)
sources of entropic gravity and matter fields.

Then, in sec. \ref{s5}, we consider the AFDM for constructing cosmological
solutions for entropic quasiperiodic flow and MGTs. We emphasize that there
are certain nonlinear symmetries relate possible classes of generating
functions and (effective) sources all encoding entropic, quasiperiodic,
pattern forming, space and time quasicrystal, solitonic and other type
structures. There are studied cosmological configurations generated by
entropic quasiperiodic sources, nonstationary generating functions,
cosmological metrics evolving in (off-) diagonal elastic and/or
quasiperiodic media.

Finally, we conclude our work and discuss certain perspectives of the theory
of G. Perelman and E. Verlinde entropic geometric flow and emergent gravity
theories in Sec. \ref{s6}.

\section{Spacetime 2+2 \& 3+1 fibrations with elastic and quasiperiodic
structures}

\label{s2} In this section, we summarize necessary results on the geometry
of Lorentz manifolds enabled with nonholonomic (i.e non-integrable,
equivalently, anholonomic) distributions defining double 2+2 and 3+1
fibrations. There are developed nonholonomic geometric methods which are
important for elaborating theories of relativistic Ricci flows and possible
applications in modern cosmology and astrophysics, see details in \cite%
{ruchin13,gheorghiu16,rajpoot17}. As explicit examples, we shall consider
nonholonomic distributions modelling elastic and/or quaisperiodic space and
time structures (for instance, quasicrystal or solitonic like
configurations) \cite%
{verlinde10,verlinde16,hossenfelder17,vacaru18tc,bubuianu17,amaral17,aschheim18}%
. It should be noted that the 2+2 nonholonomic splitting is important for
proofs of general decoupling and integration properties of the relativistic
geometric and entropic flow evolution, nonholonomic Ricci soliton and
(entropic modified) Einstein equations, see section \ref{s4}. Additional 3+1
decompositions adapted to 2+2 splitting will be used for defining and
computing entropic and thermodynamic like values for various classes of
solutions of physically important systems of nonlinear partial differential
equations, PDEs, see section \ref{s5}.

\subsection{Nonlinear connections with 2+2 splitting of Lorentz manifolds}

Let us consider a four dimensional, 4-d, Lorentzian manifold $V,$ $\dim V=4,$
with local pseudo-Euclidean signature $(+++-)$ for a metric field $\mathbf{g}%
=(h\mathbf{g},v\mathbf{g}).$ The conventional horizontal, h, and vertical,
v, nonholonomic decomposition is defined by a nonlinear connection,
N-connection, structure $\mathbf{N.}$ Such a geometric object can be always
introduced as a Whitney sum
\begin{equation}
\mathbf{N}:\ T\mathbf{V}=h\mathbf{\mathbf{V\oplus }}v\mathbf{V},
\label{whitney}
\end{equation}%
where $T\mathbf{V}$ is the tangent bundle on $\mathbf{V}$. The concept of
nonholonomic manifold is used for a manifold enabled with a nonholonomic
distribution. In this work, this refers to a Lorentz spacetime $\mathbf{V}%
:=(V,\mathbf{N})$ enabled with a N-connection structure of type (\ref%
{whitney}). In local coordinates, $\mathbf{N}=N_{i}^{a}(u)dx^{i}\otimes
\partial _{a},$ where $N_{i}^{a}$ are N--connection coefficients\footnote{%
We can parameterize the local coordinates in the form $u^{\mu
}=(x^{i},y^{a}),$ (in brief, $u=(x,y)$), where indices respectively take
values $i,j,...=1,2$ and $a,b,...=3,4,$ considering that $u^{4}=y^{4}=t$ is
the time like coordinate. The Einstein convention on summation on "up-low"
repeating indices will be applied if contrary will not be stated for some
special cases. We use boldface symbols for spaces and geometric objects
adapted to a N-connection splitting.}. Any set $\{N_{i}^{a}\}$ defines
subclasses of N-linear (co) frames which allows N-adapted diadic
decompositions of geometric and physical objects.\footnote{\label{ahr}Such
N--adapted local bases, $\mathbf{e}_{\nu }=(\mathbf{e}_{i},e_{a}),$ and
cobases, $\mathbf{e}^{\mu }=(e^{i},\mathbf{e}^{a}),$ are defined by formulas
\begin{equation*}
\mathbf{e}_{i}=\partial /\partial x^{i}-\ N_{i}^{a}(u)\partial /\partial
y^{a},\ e_{a}=\partial _{a},\mbox{ and }e^{i}=dx^{i},\mathbf{e}%
^{a}=dy^{a}+N_{i}^{a}(u)dx^{i}
\end{equation*}%
and their arbitrary frame/coordinate transforms. The term nonholonomic used
for a Lorentz manifold $\mathbf{V}$ comes from the fact that a basis
(tetrad, equivalently, vierbeind) $\mathbf{e}_{\nu }=(\mathbf{e}_{i},e_{a})$
satisfies certain relations $[\mathbf{e}_{\alpha },\mathbf{e}_{\beta }]=%
\mathbf{e}_{\alpha }\mathbf{e}_{\beta }-\mathbf{e}_{\beta }\mathbf{e}%
_{\alpha }=W_{\alpha \beta }^{\gamma }\mathbf{e}_{\gamma },$ with nontrivial
anholonomy coefficients $W_{ia}^{b}=\partial _{a}N_{i}^{b},W_{ji}^{a}=\Omega
_{ij}^{a}=\mathbf{e}_{j}\left( N_{i}^{a}\right) -\mathbf{e}_{i}(N_{j}^{a})$.
Holonomic (integrable) configurations are obtained if and only if $W_{\alpha
\beta }^{\gamma }=0.$}

On any nonholonomic manifold $\mathbf{V,}$ we can consider covariant
derivatives determined by affine (linear) connections which are, or not,
adapted to a N--connection structure. A distinguished connection, \textit{%
d--connection, } is a linear connection $\mathbf{D}=(h\mathbf{D},v\mathbf{D}%
) $ which preserves under parallel transport a h-v-decomposition (\ref%
{whitney}).\footnote{%
In general, a linear connection $D$ is not adapted to a prescribed
N-connection structure, i.e. it is not a d--connection. In such a case, one
should be not used a boldface symbols for respective geometric objects
determined by $D.$} For any d--connection $\mathbf{D,}$ we can define and
compute in standard form the d--torsion, $\mathbf{T,}$ the nonmetricity, $%
\mathbf{Q},$ and the d--curvature, $\mathbf{R},$ tensors
\begin{equation*}
\mathbf{T}(\mathbf{X,Y}):=\mathbf{D}_{\mathbf{X}}\mathbf{Y}-\mathbf{D}_{%
\mathbf{Y}}\mathbf{X}-[\mathbf{X,Y}],\mathbf{Q}(\mathbf{X}):=\mathbf{D}_{%
\mathbf{X}}\mathbf{g,} \ \mathbf{R}(\mathbf{X,Y}):=\mathbf{D}_{\mathbf{X}}%
\mathbf{D}_{\mathbf{Y}}-\mathbf{D}_{\mathbf{Y}}\mathbf{D}_{\mathbf{X}}-%
\mathbf{D}_{\mathbf{[X,Y]}},
\end{equation*}%
where $\mathbf{X}$ and$\mathbf{Y}$ are vector fields (i.e. d-vectors) on $T%
\mathbf{V.}$\footnote{%
We can compute in N--adapted form the coefficients of any d--connection $%
\mathbf{D}=\{\mathbf{\Gamma }_{\ \alpha \beta
}^{\gamma}=(L_{jk}^{i},L_{bk}^{a},C_{jc}^{i},C_{bc}^{a})\}.$ The
coefficients of torsion, nonmetricity and curvature d--tensors are
parameterized by $h$- and $v$--indices, respectively, $\mathcal{T} =\{%
\mathbf{T}_{\ \alpha \beta }^{\gamma }=\left( T_{\ jk}^{i},T_{\ ja}^{i},T_{\
ji}^{a},T_{\ bi}^{a},T_{\ bc}^{a}\right) \},\ \mathcal{Q}=\mathbf{\{Q}_{\
\alpha \beta }^{\gamma }\}, \mathcal{R} = \mathbf{\{R}_{\ \beta \gamma
\delta }^{\alpha }=\left( R_{\ hjk}^{i},R_{\ bjk}^{a},R_{\ hja}^{i},R_{\
bja}^{c},R_{\ hba}^{i},R_{\ bea}^{c}\right) \}$, when the coefficients
formulas for such values determined by using $\mathbf{\Gamma }_{\ \alpha
\beta }^{\gamma }$ and their partial derivatives.}

Any metric tensor $\mathbf{g}=(h\mathbf{g},v\mathbf{g}),$ on a nonholonomic $%
\mathbf{V}$ can be written as a distinguished tensor, d--tensor (d--metric),
with respective splitting into h- and v-indices,
\begin{equation}
\mathbf{g}=g_{\alpha }(u)\mathbf{e}^{\alpha }\otimes \mathbf{e}^{\beta
}=g_{i}(x)dx^{i}\otimes dx^{i}+g_{a}(x,y)\mathbf{e}^{a}\otimes \mathbf{e}%
^{a},  \label{dm}
\end{equation}%
where the nonholonomic dual frame structure $\mathbf{e}^{\alpha }$ is chosen
in a form when the matrix of metric coefficients $g_{\alpha \beta }$ is
considered in diagonal form for $g_{\alpha }:=g_{\alpha \alpha
},g_{i}:=g_{ii}$ and $g_{a}:=g_{aa}.$ With respect to a dual local
coordinate basis $du^{\alpha }$ the same metric field is expressed
\begin{equation}
\mathbf{g}=\underline{g}_{\alpha \beta }du^{\alpha }\otimes du^{\beta },%
\mbox{\ where \ }\underline{g}_{\alpha \beta }=\left[
\begin{array}{cc}
g_{ij}+N_{i}^{a}N_{j}^{b}g_{ab} & N_{j}^{e}g_{ae} \\
N_{i}^{e}g_{be} & g_{ab}%
\end{array}%
\right] .  \label{offd}
\end{equation}%
Using frame transforms (in general, not N--adapted), we can transform any
metric into a d--metric (\ref{dm}) an off-diagonal form with
N--coefficients. For nontrivial anholonomy coefficients, such a metric is
generic off-diagonal.

For our geometric constructions, there are two important linear connections
determined by the same metric structure: \vskip2pt {\
\begin{equation}
\mathbf{g}\rightarrow \left\{
\begin{array}{ccccc}
\nabla : &  & \nabla \mathbf{g}=0;\ ^{\nabla }\mathbf{T}=0, &  &
\mbox{ the
Levi--Civita, LC, connection;} \\
\mathbf{D}: &  & \mathbf{D}\ \mathbf{g}=0;\ h\mathbf{T}=0,\ v\mathbf{T}=0. &
& \mbox{ the canonical
d--connection.}%
\end{array}%
\right.  \label{lcconcdcon}
\end{equation}%
Here we note that if we prescribe a N-connection structure }$\mathbf{N},$ we
can define a canonical d-connection $\mathbf{D}$ and compute certain
nontrivial torsion coefficients $hv\mathbf{T}$ completely defined by certain
off-diagonal coefficients containing $N_{i}^{a}(u)$ in (\ref{offd}) and/or
nontrivial anholonomy coefficients $W_{\alpha \beta }^{\gamma },$ see
footnote \ref{ahr}. Of course, we can introduce an infinite number of metric
compatible d-connections but not all such connections allow to decouple
physically importants systems of nonlinear PDEs (in our case, for
nonholonomic geometric flows and modified, or Einstein, gravity). A $\mathbf{%
D}$ (\ref{lcconcdcon}) allows us to prove the decoupling and integration
properties of such equations in section \ref{s4}.

The LC--connection $\nabla $ (\ref{lcconcdcon}) can be defined uniquely by a
metric $\mathbf{g}$ without any N--connection structure but $\nabla $ can be
distorted always to a necessary type d--connection allowing a general
decoupling and integrability of certain important physically important
systems of nonlinear PDEs. In our previous works \cite%
{ruchin13,gheorghiu16,rajpoot17} (see there necessary geometric details and
\cite{vacaru18tc,bubuianu17,amaral17,aschheim18}\ for applications of the
AFDM), we used a "hat" symbol (like $\widehat{\mathbf{D}}$) for the
canonical d-connection in (\ref{lcconcdcon}). In this paper, we shall work
only with $\nabla $ and $\mathbf{D=}$ $\widehat{\mathbf{D}}$ and omit "hats"
on respective geometric objects. We note that all constructions performed
for $\nabla $ and $\mathbf{D}$ are related by a distortion relation, $%
\mathbf{D[g,N]}=\nabla \mathbf{[g,N]}+\mathbf{Z[g,N],}$ where $\mathbf{Z}$
is the distortion tensor determined in standard algebraic form by the
torsion tensor $\mathbf{T;}$ all values are completely defined by the metric
tensor $\mathbf{g}$ adapted to $\mathbf{N.}$\footnote{%
The values $h\mathbf{T}$ and $\ v\mathbf{T}$ are respective torsion
components which vanish on conventional h- and v--subspaces, but there are
nontrivial components $hv\mathbf{T}$ defined by certain anholonomy
(equivalently, nonholonomic/ non-integrable) relations. Such a d-torsion is
induced by nonholonomic configurations.}

The Ricci tensors of $\mathbf{D}$ and $\nabla $ are defined and computed in
standard forms for different linear connection structures but defined by the
same metric tensor by contracting respective indices. We denote them,
respectively, $\mathbf{R}ic=\{\mathbf{R}_{\ \alpha \beta }:=\mathbf{R}_{\
\alpha \beta \gamma }^{\gamma }\}$ and $Ric=\{R_{\ \alpha \beta }:=R_{\
\alpha \beta \gamma }^{\gamma }\}.$ Any (pseudo) Riemannian geometry can be
equivalently described by both geometric data $\left( \mathbf{g,\nabla }%
\right) $ and $(\mathbf{g,N,D}),$ when the canonical distortion relations $\
\mathbf{R}=\ ^{\nabla }\mathbf{R}\mathcal{+}\ ^{\nabla }\mathbf{Z}$ and $\
\mathbf{R}ic=Ric+Zic,$ with respective distortion d-tensors $\ ^{\nabla }%
\mathbf{Z}$ and $Zic,$ are computed for the canonical distortion relations $%
\mathbf{D}=\nabla +\mathbf{Z},$ see details in \cite%
{bubuianu18,vacaru09,ruchin13,gheorghiu16,rajpoot17} (in those works, there
are used different systems of notations).

Using N-adapted coefficients of the canonical Ricci d-tensor,
\begin{equation}
\mathbf{R}_{\alpha \beta }=\{\mathbf{R}_{ij}:=\mathbf{R}_{\ ijk}^{k},\
\mathbf{R}_{ia}:=-\mathbf{R}_{\ ika}^{k},\ \mathbf{R}_{ai}:=\mathbf{R}_{\
aib}^{b},\ \mathbf{R}_{ab}:=\mathbf{R}_{\ abc}^{c}\},  \label{driccic}
\end{equation}%
we can compute the scalar of canonical d--curvature, $\ ^{s}\mathbf{R}:=%
\mathbf{g}^{\alpha \beta }\mathbf{R}_{\alpha \beta }=g^{ij}\mathbf{R}%
_{ij}+g^{ab}\mathbf{R}_{ab}.$ This geometric object is different from the
scalar curvature of the LC-connection, $\ R:=\mathbf{g}^{\alpha
\beta}R_{\alpha \beta }.$

Using $\nabla ,$ the Einstein equations in GR are written in standard form,
\begin{equation}
R_{\alpha \beta }-\frac{1}{2}g_{\alpha \beta }R=\varkappa \ ^{m}T_{\alpha
\beta }.  \label{einsteq}
\end{equation}
In these formulas, $\ ^{m}T_{\alpha \beta }$ is the energy--momentum tensor
of matter fields $\ ^{A}\varphi $ determined by a general Lagrangian $\ ^{m}%
\mathcal{L}(\mathbf{g,}\nabla ,\ \ ^{A}\varphi ),$ where $\varkappa $ is the
gravitational coupling constant for GR.\footnote{%
We use abstract left labels $A$ and $m$ in order to distinguish the values
from similar notations of pure geometric objects, for instance, $\mathbf{T}%
_{\ \alpha \beta }^{\gamma }.$}

We can define nonholonomic gravitational field equations using the Ricci
d-tensor (\ref{driccic}) for a canonical d-connection $\mathbf{D}$
\begin{equation}
\mathbf{R}_{\alpha \beta }=\mathbf{\Upsilon }_{\alpha \beta }.
\label{deinst}
\end{equation}%
Such equations are equivalent to (\ref{einsteq}) if there are imposed
additional nonholonomic constraints, or found some smooth limits, for
extracting LC--configurations, $\mathbf{D}_{\mid \widehat{\mathcal{T}}%
=0}=\nabla ,$ for instance, of type
\begin{equation}
\mathbf{T}_{\ \alpha \beta }^{\gamma }=0.  \label{zerot}
\end{equation}%
In (\ref{deinst}), a matter fields source $\mathbf{\Upsilon }_{\mu \nu }$
can be constructed using a N--adapted variational calculus for $\ ^{m}%
\mathcal{L}(\mathbf{g,}\widehat{\mathbf{D}},\ \ ^{A}\varphi ),$ when $%
\mathbf{\Upsilon }_{\mu \nu }=\varkappa (\ ^{m}\mathbf{T}_{\mu \nu }-\frac{1%
}{2}\mathbf{g}_{\mu \nu }\ ^{m}\mathbf{T})\rightarrow \varkappa (\
^{m}T_{\mu \nu }-\frac{1}{2}\mathbf{g}_{\mu \nu }\ ^{m}T)$ for [coefficients
of $\ \mathbf{D}]$ $\rightarrow $ [coefficients of $\nabla$] even, in
general, $\mathbf{D}\neq \nabla .$ In such formulas, we consider $^{m}%
\mathbf{T}=\mathbf{g}^{\mu \nu }\ ^{m}\mathbf{T}_{\mu \nu }$ for
\begin{equation}
\ ^{m}\mathbf{T}_{\alpha \beta }:=-\frac{2}{\sqrt{|\mathbf{g}_{\mu \nu }|}}%
\frac{\delta (\sqrt{|\mathbf{g}_{\mu \nu }|}\ \ ^{m}\mathcal{L})}{\delta
\mathbf{g}^{\alpha \beta }}.  \label{ematter}
\end{equation}

We note that any (pseudo) Riemannian geometry and gravity theory, and
various metric-affine modifications (for instance, $F(R)$-modified theories
\cite{nojiri17,bubuianu18}), can be formulated equivalently using geometric
data $(\mathbf{g,\nabla )}$ and/or $(\mathbf{g},\mathbf{D}).$ There is an
important motivation to use nonholonomic variables of type $(\mathbf{g},%
\mathbf{D})$ because that they allow to decouple and integrate in general
form various modified and standard Einstein equation. Such solutions can be
with generic off-diagonal metrics and coefficients depending on all
spacetime coordinates \cite{v05,v09,v09a}. A recent review of the so-called
anholonomic frame deformation method, AFDM, of constructing exact solutions
in GR and MGTs, geometric flow theory, and applications in modern cosmology
and astrophysics can be found in \cite{bubuianu18}. In this work, we shall
develop the AFDM for constructing exact solutions in entropic geometric flow
and gravity theories.

\subsection{Nonholonomic 3+1 splitting adapted to 2+2 structures}

\label{ss31sp}We outline some basic concepts on the geometry of 3+1
foliations of a nonholonomic Lorentzian manifold $(\mathbf{V,g,N)}$ of
signature $(+++-)$ into a family of non-intersecting space like 3-d
hypersurfaces $\Xi _{t}$ parameterized by a "time function", $t(u^{\alpha}),
$ stated as a scalar field as described as follows. Such spacetime
decompositions are useful for elaborating various thermodynamic, locally
anisotropic kinetic \cite{vacaru00ap} and geometric evolution or
hydrodynamic flow models \cite{ruchin13} when a conventional splitting into
time and space like coordinates is necessary. This allows definition of
physical important values (for instance, entropy, effective energy etc.) and
deriving fundamental geometric evolution equations. In our approach, we
generalize the well--known geometric 3+1 formalism in GR (see, for instance,
\cite{misner}) to the case of nonholonomic manifolds \cite%
{ruchin13,gheorghiu16,rajpoot17}.

For a 3--d manifold $\ _{\shortmid }\Xi ,$ we consider an one-to-one image
to a hypersurface $\Xi =\zeta (\ _{\shortmid }\Xi )\subset \mathbf{V}$
constructed as an homeomorphism with both continuous maps $\zeta $ and $%
\zeta ^{-1},$ when $\Xi $ does not intersect itself. Left "up" or "low"
labels by a vertical bar "$\ _{\shortmid }\ $" will be used in order to
emphasize that certain geometric objects refer to 3--d manifolds /
hypersurfaces. Such a 3-d space is supposed to be locally defined as a set
of points for which a scalar field $t$ on $\mathbf{V}$ is constant (for
instance, i.e. $t(p)=0$ for any point $p\in \Xi ).$ It is assumed also that $%
t$ spans the real line $\mathbb{R}$ and that any $\Xi $ is a connected
submanifold of $\mathbf{V}$ with the topology of $\mathbb{R}^{3}.$\footnote{%
We can label local coordinates for a 3+1 splitting in $u^{\alpha }=(x^{%
\grave{\imath}},t),$ where indices $\alpha ,\beta ,...=1,2,3,4$ and $\grave{%
\imath},\grave{j},...=1,2,3$ are related to a 2+2 splitting as in previous
subsection (in brief, we shall write $u=(\breve{u},t)).$ The continuous maps
$\zeta $ can be parameterized to "carry along" curves/ vectors in $\
_{\shortmid }\Xi $ to curves / vectors in $\mathbf{V,}$ for $\zeta :(x^{%
\grave{\imath}})\longrightarrow (x^{\grave{\imath}},0).$ This way, it is
possible to define and relate respective local bases $\partial _{\grave{%
\imath}}:=\partial /\partial x^{\grave{\imath}}\in T \mathbf{(\ _{\shortmid
}\Xi )}$ and $\partial _{\alpha }:=\partial /\partial u^{\alpha }\in T%
\mathbf{V.}$ The coefficients of 3--vectors and 4--vectors are expressed
correspondingly,$\ _{\shortmid }\mathbf{a}=a^{\grave{\imath}}\partial _{%
\grave{\imath}}$ and $\mathbf{\ a}=a^{\alpha }\partial _{\alpha} $ (we shall
use also capital letters, for instance, $\ _{\shortmid }\mathbf{A}=A^{\grave{%
\imath}}\partial _{\grave{\imath}}$ and $\mathbf{\ A}=A^{\alpha }\partial
_{\alpha }).$ Similar formulas are considered for dual forms to vectors,
1--forms, when the dual bases $\mathbf{d}x^{\grave{\imath}}\in T^{\ast }%
\mathbf{(\mathbf{\ }_{\shortmid }\Xi )}$ and $du^{\alpha }\in T^{\ast }%
\mathbf{V.}$ The 1--forms will be parameterized for respective 3 and 4
dimensions, $\ _{\shortmid }\tilde{\mathbf{A}}=A_{\grave{\imath}}\mathbf{d}%
x^{\grave{\imath}}$ and $\mathbf{\ \tilde{A}}=A_{\alpha }du^{\alpha }.$ We
shall omit the left/up label by a tilde $\sim $ (writing $\mathbf{\ }%
_{\shortmid }\mathbf{A}$ and $\mathbf{A)}$ if that will not result in
ambiguities.}

It should be noted that any 2+2 splitting by a nonholonomic distribution $%
\mathbf{N}$ (\ref{whitney}) induces a N--connection structure for a
hypersurface $\Xi ,$ i.e. an induced N-connections $\ _{\shortmid }\mathbf{N}%
:T\ _{\shortmid }\Xi =h\ _{\shortmid }\Xi \oplus v\ _{\shortmid }\Xi \mathbf{%
.}$ Using the coefficients of such an induced N-connection, any induced
3--metric tensor $\mathbf{q}$ can be written in N--adapted frames as a
d--tensor (d--metric) in the form
\begin{eqnarray*}
\mathbf{q} &=&(h\mathbf{q},v\mathbf{q})=q_{\grave{\imath}}(u)\mathbf{e}^{%
\grave{\imath}}\otimes \mathbf{e}^{\grave{\imath}}=q_{i}(x^{k})dx^{i}\otimes
dx^{i}+q_{3}(x^{k},y^{3})\ _{\shortmid }\mathbf{e}^{3}\otimes \ _{\shortmid }%
\mathbf{e}^{3}, \\
&&\mbox{for}\ _{\shortmid }\mathbf{e}^{3}=du^{3}+\ _{\shortmid
}N_{i}^{3}(u)dx^{i},
\end{eqnarray*}%
where $\ _{\shortmid }N_{i}^{3}(u)$ can be identified with $N_{i}^{3}(u)$
choosing common frame and coordinate systems for $\Xi \subset \mathbf{V}$.
We can extend naturally such a 3-d metric $\mathbf{q}$ to a 4-d d--metric $%
\mathbf{g}$ (\ref{dm}) re-parameterized in a form adapted both to 2+2 and
3+1 nonholonomic splitting,
\begin{eqnarray}
\mathbf{g} &=&(h\mathbf{g},v\mathbf{g})=\breve{g}_{\grave{\imath}\grave{j}}%
\mathbf{e}^{\grave{\imath}}\otimes \mathbf{e}^{\grave{j}}+g_{4}\mathbf{e}%
^{4}\otimes \mathbf{e}^{4}=q_{\grave{\imath}}(u)\mathbf{e}^{\grave{\imath}%
}\otimes \mathbf{e}^{\grave{\imath}}-\breve{N}^{2}\mathbf{e}^{4}\otimes
\mathbf{e}^{4},  \label{offdublespl} \\
\mathbf{\ e}^{3} &=&\ _{\shortmid }\mathbf{e}^{3}=du^{3}+_{\shortmid
}N_{i}^{3}(u)dx^{i},\mathbf{\ e}^{4}=\delta t=dt+N_{i}^{4}(u)dx^{i}.  \notag
\end{eqnarray}%
For $\mathbf{g}$ (\ref{offdublespl}),\ the lapse function $\breve{N}(u)>0$
is defined as a positive scalar field which ensues that the d--vector $%
\mathbf{n}$ is a unite one. An "inverse hat" symbol is used in order to
distinguish such a symbol from $N$ is used traditionally in literature on GR
\cite{misner}. Here we note that in another turn, the symbol $N_{i}^{a}$ is
used traditionally for the N--connection and this also motivates a new
symbol $\breve{N}$.

We note that for any quadratic line element $ds^{2}=g_{\alpha \beta
}du^{\alpha }du^{\beta }$ of a metric tensor $\mathbf{g}$ there are such
frame transforms to parameterizations when $\breve{g}_{\grave{\imath}\grave{j%
}}=q_{\grave{\imath}\grave{j}}=$ $g_{\alpha \beta }e_{~\grave{\imath}%
}^{\alpha }e_{~\grave{j}}^{\beta }$ is just the induced metric on $\Xi _{t}$%
. In result, the determinants of 4-d and 3-d metrics are computed $\sqrt{|g|}%
=\breve{N}\sqrt{|\breve{g}|}=\breve{N}\sqrt{|q|}.$ Using certain coordinates
$(x^{\grave{\imath}},t)$ being N-adapted on respective hypersurfaces, the
time partial derivatives are computed $\pounds _{t}q=\partial _{t}q=q^{\ast
} $ and the spacial derivatives are computed $q_{,\ ^{\grave{\imath}}}:=e_{\
^{^{\grave{\imath}}}}^{\alpha }$ $q_{,\alpha }.$

There are two types of induced linear connections completely determined by
an induced 3--d hypersurface metric $\mathbf{q,}$ \vskip2pt

\begin{equation}
\mathbf{q}\rightarrow \left\{
\begin{array}{cccc}
\ _{\shortmid }\nabla : &  & \mathbf{\ }_{\shortmid }\nabla \mathbf{q}=0;\
\mathbf{\ }_{\shortmid }^{\nabla }\mathcal{T}=0, & \mbox{LC--connection} \\
\ _{\shortmid }\mathbf{D}: &  & \ _{\shortmid }\mathbf{D}\ \mathbf{q}=0;\ h\
_{\shortmid }\mathcal{T}=0,\ v\ _{\shortmid }\mathcal{T}=0, & %
\mbox{canonical d--connection.}%
\end{array}%
\right.  \label{hlccdc}
\end{equation}%
\vskip2pt

Such formulas are related to 4--d similar ones (\ref{lcconcdcon}). Both
linear connections, $\ _{\shortmid }\nabla $ and $\ _{\shortmid }\mathbf{D},$
are subjected also to a distortion relation $\ _{\shortmid }\mathbf{D[q,\
_{\shortmid }N]}=\ _{\shortmid}\nabla \lbrack \mathbf{q}]+\ _{\shortmid}%
\mathbf{Z}[\ _{\shortmid }\mathcal{T}(q,\ _{\shortmid }\mathbf{N})].$

For 3-d configurations, we can compute the N--adapted coefficient formulas
for nonholonomically induced torsion structure $\mathbf{\ _{\shortmid }T}%
=\{\ _{\shortmid }\mathbf{T}_{\ \grave{j}\grave{k}}^{\grave{\imath}}\},$ \
determined by $\ _{\shortmid }\mathbf{D,}$ and for the Riemannian tensors $\
_{\shortmid }R=\{\ \ _{\shortmid }R_{\ \grave{j}\grave{k}\grave{l}}^{\grave{%
\imath}}\}$ and $\mathbf{\mathbf{\ }_{\shortmid }R}=\mathbf{\{_{\shortmid }R}%
_{\ \grave{j}\grave{k}\grave{l}}^{\grave{\imath}}\},$ determined
respectively by $\ _{\shortmid }\nabla $ and $\ _{\shortmid }\mathbf{D.}$
Using 3--d subsects of coefficient formulas, we can compute respective
N--adapted hypersurface coefficients of the Ricci d--tensor, $\ _{\shortmid }%
\mathbf{R}_{\ \grave{j}\grave{k}},$ and the Einstein d--tensor, $\
_{\shortmid }\mathbf{E}_{\ \grave{j}\grave{k}}.$ Contracting indices, we
obtain the Gaussian curvature, $\ _{\shortmid }R=q^{\grave{j}\grave{k}}\
_{\shortmid }R_{\grave{j}\grave{k}}$, and the Gaussian canonical curvature, $%
\ _{\shortmid }^{s}R=q^{\grave{j}\grave{k}}\ _{\shortmid }\mathbf{R}_{\grave{%
j}\grave{k}},$ of $(\Xi ,\mathbf{q,\ _{\shortmid }N}).$ It should be noted
that all this types of N-adapted and not N-adapted geometric objects can be
defined in abstract form which do not depend on the type of embedding of a
nonholonomic 3-d manifold $(\Xi ,\mathbf{q,\ _{\shortmid }N})$ into a 4-d
one $(\mathbf{V},\mathbf{g,N}).$

\subsection{Quasiperiodic space \& time QC configurations}

Let us consider two examples of space and time quasiperiodic structures
defined in a curved spacetime following our works on quasicrystal, QC,
models in modern cosmology \cite{vacaru18tc,bubuianu18,amaral17}
(alternative models are studied in \cite{ghosh}). Our approach was
elaborated following F. Wilczek and co-authors works in condensed matter
physics \cite{shapere12,wilczek12,wilczek13a,shapere17}. As a toy model, we
consider one dimensional, 1-d, time quasicrystals, TQCs with time structure
equations generalizing those introduced in \cite{shapere17}). Then we
introduce some important formulas on three dimensional, 3-d, QC structures -
in general, such configurations are called space-time quasicrystal
structures, STQC, and studied in \cite{vacaru18tc}. In this work we use a
different system of notation for partial derivatives when, for instance, $%
\partial q/\partial x^{i}=\partial _{i}q,$ $\partial q/\partial
y^{3}=\partial _{3}q=q^{\ast },$ and $\partial q/\partial y^{4}=\partial
_{4}q=\partial _{t}q=q^{\star },$ for a function $q(x^{i},y^{3},t).$

\subsubsection{1-d relativistic time QC structures}

We consider a scalar field $\varsigma (x^{i},y^{a})$ on a space-time $(%
\mathbf{V,g,N)}$ and respective Lagrange density
\begin{equation}
\acute{L}(\varsigma )=\frac{1}{48}(\mathbf{g}^{\alpha \beta }(\mathbf{e}%
_{\alpha }\varsigma )(\mathbf{e}_{\beta }\varsigma ))^{2}-\frac{1}{4}\mathbf{%
g}^{\alpha \beta }(\mathbf{e}_{\alpha }\varsigma )(\mathbf{e}_{\beta
}\varsigma )-\acute{V}(\varsigma ).  \label{1tcqc}
\end{equation}%
In this formula, $\acute{V}(\varsigma )$ is a nonlinear potential and $%
\mathbf{e}_{\alpha }$ are N-adapted partial derivatives. Corresponding
N-adapted variational motion equations are $\lbrack \frac{1}{2}\mathbf{g}%
^{\alpha \beta }(\mathbf{e}_{\alpha }\varsigma)(\mathbf{e}_{\beta }\varsigma
)-1](\mathbf{D}^{\gamma }\mathbf{D}_{\gamma }\varsigma )=2\frac{\partial
\acute{V}}{\partial \varsigma }$. The field $\varsigma $ defines a 1-d time
QC structure, 1-TQC, if it is a solution of these motion equations.\footnote{%
For non-relativistic limits with $g_{\alpha \beta }=[1,1,1,-1]$ and $%
\varsigma \rightarrow \varsigma (t),$ $\acute{L}\rightarrow \frac{1}{12}%
(\varsigma ^{\bullet })^{4} - \frac{1}{2}(\varsigma ^{\bullet })^{2}-\acute{V%
}(\varsigma ),$ which leads to an effective energy $E=\frac{1}{4}[(\varsigma
^{\bullet })^{2}-1]^{2}+\acute{V}(\varsigma )-\frac{1}{4}$ and motion
equations $[(\varsigma ^{\bullet })^{2}-1]\varsigma ^{\bullet \bullet }=-
\frac{\partial \acute{V}}{\partial \varsigma }$ \ introduced in \cite%
{shapere17}. The Lagrange density (\ref{1tcqc}) provides a generalization
for 1-TQCs modeled on a curved spacetime which can be also modeled in
entropic gravity theories.}

\subsubsection{3-d QC structures on curved spaces}

QC structures and analogous dynamic phase field crystal models can be
elaborated as flow evolution theories on real parameter ${\tau }$ (in next
Section, this parameter will be identified with a geometric flows one). Such
a QC structure can be defined by a generating function $\ \overline{q}=%
\overline{q}(x^{i},y^{3},\tau )$ subjected to the condition that it is a
solution of an evolution equation with conserved dynamics,
\begin{equation}
\frac{\partial \overline{b}}{\partial \tau }=\ \ _{\shortmid }\widehat{%
\Delta }\left[ \frac{\delta \overline{F}}{\delta \overline{b}}\right] =-\ \
_{\shortmid }\widehat{\Delta }(\Theta \overline{b}+Q\overline{b}^{2}-%
\overline{b}^{3}).  \label{evoleq}
\end{equation}%
Such evolution is considered on 3-d spacelike hypersurface $\Xi _{t}$ when
the canonically nonholonomically deformed hypersurface Laplace operator $\ \
_{\shortmid }\widehat{\Delta }:=(\ _{\shortmid }\mathbf{D})^{2}=q^{\grave{%
\imath}\grave{j}}\ _{\shortmid }\mathbf{D}_{\grave{\imath}}\ _{\shortmid }%
\mathbf{D}_{\grave{j}},$ where indices rung values $\grave{\imath},\grave{j}%
,...1,2,3.$ This operator is a distortion of $\ \ _{\shortmid }\Delta :=(\
_{\shortmid }\nabla )^{2}$ constructed in 3-d Riemannian geometry, see
previous subsection. The functional $\overline{F}$ in (\ref{evoleq}) is
characterized by an effective free energy
\begin{equation*}
\overline{F}[\overline{q}]=\int \left[ -\frac{1}{2}\overline{b}\Theta
\overline{b}-\frac{Q}{3}\overline{b}^{3}+\frac{1}{4}\overline{b}^{4}\right]
\sqrt{q}dx^{1}dx^{2}\delta y^{3},
\end{equation*}
where $q=\det |q_{\grave{\imath}\grave{j}}|,\delta y^{3}=\mathbf{e}^{3}$ and
the operators $\Theta $ and $Q$ are defined and explained in \cite%
{bubuianu18,vacaru18tc}. Such nonlinear interactions are stabilized by the
cubic term with $Q$ and the second order resonant interactions are varied by
setting observable values of such constants (they are different for
cosmological models, in astrophysics or condensed matter physics). The
average value $<\overline{b}>$ is conserved for any fixed time variable $t$
and/or evolution parameter $\tau _{0}.$ We can fix $<\overline{b}>_{|\tau
=\tau _{0}}=0$ when other values are accommodated by redefining values $%
\Theta $ and $Q.$

\subsection{Distributions defining spacetime elastic configurations}

\label{sselastic}In letter \cite{partner1}, we shown that models of entropic
gravity can be derived from nonholonomic modifications of the W-functional
when the (modified) Einstein equations are equivalent to certain
nonholonomic Ricci soliton equations. Here, we shall study the conditions
when entropic elastic scenarios can be modelled as nonholonomic Ricci
solitons in subsection \ref{ssentrg}. We shall consider certain examples of
nonholonomic distributions and related Lagrange densities on a Lorentz
manifold $\mathbf{V}$ which are used in entropic gravity theories \cite%
{verlinde10,verlinde16,hossenfelder17,partner1}. Using such geometric
constructions, we shall elaborate in next section on elastic flow evolution
models and their self-similar nonholonomic Ricci soliton configurations.
There are three important values:
\begin{eqnarray*}
\varepsilon _{\alpha \beta } &=&\mathbf{D}_{\alpha }\mathbf{u}_{\beta }-%
\mathbf{D}_{\beta }\mathbf{u}_{\alpha }\mbox{\ - the elastic strain tensor }%
;\phi =u/\sqrt{|\Lambda |}\mbox{ - a dimensionless scalar }; \\
\chi &=&\alpha (\mathbf{D}_{\mu }\mathbf{u}^{\mu })(\mathbf{D}_{\nu }\mathbf{%
u}^{\nu })+\beta (\mathbf{D}_{\mu }\mathbf{u}_{\nu })(\mathbf{D}^{\mu }%
\mathbf{u}^{\nu })+\gamma (\mathbf{D}_{\mu }\mathbf{u}_{\nu })(\mathbf{D}%
^{\nu }\mathbf{u}^{\mu })\mbox{ - a general kinetic term for }\mathbf{u}%
^{\mu }.
\end{eqnarray*}%
These geometric/physical objects are determined by a conventional
displacement vector field $\mathbf{u}^{\alpha },$ cosmological constant $%
\Lambda $ and some constants $\alpha ,\beta ,\gamma ;$ there are used short
hand notations: $u:=\sqrt{|\mathbf{u}_{\alpha }\mathbf{u}^{\alpha }|}%
,\varepsilon =\varepsilon _{\beta }^{\beta },$ and $\mathbf{n}^{\alpha }:=%
\mathbf{u}^{\alpha }/u.$

On $\mathbf{V,}$ there are considered nonholonomic distributions for
corresponding total, effective gravitational, usual matter, interaction and
kinetic terms of Lagrangians postulated in the form
\begin{eqnarray}
\ ^{tot}\mathcal{L} &=&\ ^{g}\mathcal{L+}\ ^{m}\mathcal{L+}\ ^{int}\mathcal{%
L+}\ ^{\chi }\mathcal{L},\mbox{ for }  \label{lagrs} \\
\ ^{g}\mathcal{L} &=&M_{P}^{2}F(\ ^{s}R),\ ^{int}\mathcal{L}=-\sqrt{|\Lambda
|}\ ^{m}\mathbf{T}_{\mu \nu }\mathbf{u}^{\mu }\mathbf{u}^{\nu }/u,\ ^{\chi }%
\mathcal{L=}M_{P}^{2}|\Lambda |(\chi ^{3/2}+|\Lambda ||u[\varsigma ,%
\overline{b}]|^{2z}).  \notag
\end{eqnarray}%
In these formulas, the Plank gravitational mass is denoted $M_{P}$ and the
gravitational Lagrangian $\ ^{g}\mathcal{L}$ is taken as in modified gravity
\cite{nojiri17,capozziello,hossain15,bubuianu18}. We can fix $z=1$ if we
search for compatibility with \cite{hossenfelder17}, or $z=2$ if we search
for a limit to the standard de Sitter space solution \cite{dai17a,dai17b}
(as we use in \cite{partner1}). To model STQC structures in entropic gravity
and related geometric flow theories we can consider that the displacement
vector field $\mathbf{u}^{\alpha }[\varsigma ,\overline{b}]$ is a functional
of functions $\varsigma ,\overline{b}$ subjected to certain conditions of
type (\ref{1tcqc}) and/or (\ref{evoleq}) [in principle, we can consider
functionals for pattern forming, nonlinear wave soliton structures,
fractional and diffusion processes etc.].

The energy-momentum tensors considered in above formulas and/or derived from
respective Lagrangians in (\ref{lagrs}) and computed using variations on $%
\mathbf{g}^{\mu \nu }$ similarly to $\ ^{m}\mathbf{T}_{\mu \nu }$ (\ref%
{ematter}) (in N-adapted form, details of such computations are provided in
\cite{gheorghiu14,gheorghiu16,bubuianu17,bubuianu18}). For the full system,
the effective energy-momentum tensor is computed
\begin{eqnarray*}
\ ^{tot}\mathbf{T}_{\mu \nu } &=&(\frac{\partial F}{\partial \ ^{s}R})^{-1}\
^{m}\mathbf{T}_{\mu \nu }+\ ^{F}\mathbf{T}_{\mu \nu }+\ ^{int}\mathbf{T}%
_{\mu \nu }+\ ^{\chi }\mathbf{T}_{\mu \nu },\mbox{ where } \\
\ ^{F}\mathbf{T}_{\beta \gamma } &=&[\frac{1}{2}(F-\frac{\partial F}{%
\partial \ ^{s}R})\mathbf{g}_{\beta \gamma }-(\mathbf{g}_{\beta \gamma }%
\mathbf{D}_{\alpha }\mathbf{D}^{\alpha }-\mathbf{D}_{\beta }\mathbf{D}%
_{\gamma })\frac{\partial F}{\partial \ ^{s}R}](\frac{\partial F}{\partial \
^{s}R})^{-1}.
\end{eqnarray*}%
We can model "pure" elastic spacetime modifications of the Einstein gravity
if we fix $F(\ ^{s}R)=\ ^{s}R$ and consider restrictions to the Levi-Civita
connection $\mathbf{D=\nabla .}$ For such conditions, we obtain respective
formulas for $\ ^{int}\mathbf{T}_{\mu \nu }$ and $\ ^{\chi }\mathbf{T}_{\mu
\nu }$ which are similar to formulas (10)-(13) in \cite{dai17a}.\footnote{%
We use a system of notations which is similar (but without "hats") to \cite%
{bubuianu18,vacaru09,ruchin13,gheorghiu16,rajpoot17}; such notations are
different from those used in \cite{hossenfelder17,dai17a,dai17b}.}

In this work, the generalized (effective) source for MGT (\ref{deinst})
splits into four components,%
\begin{equation}
\ ^{tot}\mathbf{\Upsilon }_{\mu \nu }:=\varkappa \left( \ ^{tot}\mathbf{T}%
_{\mu \nu }-\frac{1}{2}\mathbf{g}_{\mu \nu }\ ^{tot}\mathbf{T}\right) =(%
\frac{\partial F}{\partial \ ^{s}R})^{-1}\ ^{m}\mathbf{\Upsilon }_{\mu \nu
}+\ ^{F}\mathbf{\Upsilon }_{\mu \nu }+\ ^{int}\mathbf{\Upsilon }_{\mu \nu
}+\ ^{\chi }\mathbf{\Upsilon }_{\mu \nu },  \label{totsourc}
\end{equation}%
where $\varkappa $ is determined in standard form by the Newton
gravitational constant $G.$ We need additional terms and parameterizations
in order to describe structure formation in modern cosmology and to model
dark energy and dark matter properties.

\section{Relativistic geometric flows and modified entropic gravity}

\label{s3}

Grigory Perelman's proof of the Poincar\'{e} conjecture \cite{perelman1} on
geometric flow evolution of 3-d Riemannian metrics \cite%
{hamilt1,hamilt2,hamilt3} provided fundamental results in geometric analysis
and topology. There were also studied possible applications in modern
mathematical and particle physics. We cite \cite%
{monogrrf1,monogrrf2,monogrrf3} for reviews of rigorous mathematical
results. For early applications, we refer to D. Friedan works \cite%
{friedan1,friedan2,friedan3} (he considered geometric evolution equations
related to renorm group equations before the Hamilton--Poincar\'{e} theory
was elaborated). Further developments and applications were performed in
\cite{verlinde00,tsey,verlinde13} and a series of works \cite%
{vacaru06,vacaru07,vacaru09,vacaru10,vacaru13,ruchin13,gheorghiu16,rajpoot17,bubuianu18a}%
, see also references therein. In those works on theories of nonholonomic/
noncommutative/ supersymmetrics, fractional, diffusion etc. geometric flows,
there were studied statistical and thermodynamic evolution models derived
from certain Lyapunov type functionals. Such F- and W-entropy functionals
are called in literature the Perelman functionals. The W-entropy has
properties of "minus entropy" of statistical thermodynamics systems. In \cite%
{partner1}, we elaborate on the idea that such a W-entropy can be considered
for formulating E. Verlinde type entropic gravity theories \cite%
{verlinde10,verlinde16}. We study self-similar configurations of
nonholonomic geometric flows resulting in entropic Ricci solitons (see
subsection \ref{ssentrg}).

We note that G. Perelman suggested in his first preprint \cite{perelman1}
that the geometric flow theory may have certain implications in black hole
physics and string theory. Nevertheless, the original theory of Ricci flows
was formulated in a non--relativistic form. To consider further
generalizations and applications in modern physics and cosmology we
elaborated on relativistic models of geometric flow theories \cite%
{ruchin13,gheorghiu16,rajpoot17}. Such constructions can be re-defined for
nonholonomic configurations modeling elastic and quasiperiodic spacetime
structure as in subsection \ref{sselastic} and allows rigorous geometric
motivations for emergent entropic theories of type \cite%
{verlinde10,verlinde16,hossenfelder17,partner1}.

The goal of this section is to study generalizations of the hypersurface 3-d
and relativistic 4-d F- and W-functionals and elaborate on respective
geometric evolution scenarios supporting the E. Verlinde entropic gravity
conjecture \cite{verlinde10,verlinde16}. Such constructions can be
considered in the framework of s a modified relativistic variant of the
Poincar\'{e}--Thurston conjecture which was proven only for certain classes
of Riemennian and K\"{a}hler metrics, see details in \cite%
{perelman1,monogrrf1,monogrrf2,monogrrf3}. For relativistic configurations,
we can only elaborate on geometric evolution of certain 3-d hypersurface
configurations subjected to the conditions that such 3-metrics can be
extended to certain classes of 4-d metric and (non) linear connection
structures derived as exact/ parametric solutions of certain nonholonomic/
entropic geometric flow equations. There will be considered also
generalizations of the Hamilton equations for the entropic flow theory. The
conditions for generating entropic modified Einstein equations as
nonholonomic Ricci solitons will be also analysed. We emphasize that in the
main part of this article there are studied relativistic geometric flow
models with a temperature like evolution parameter.

\subsection{Modified spacetime and hypersurface Perelman's functionals}

Let us consider families of nonholonomic 4-d manifolds $\mathbf{V}(\tau )$
parameterized by a positive parameter $\tau ,0\leq \tau \leq \tau _{0}$ (it
can be considered as a temperature like parameter) and enabled with a double
nonholonomic 2+2 and 3+1 splitting \cite{ruchin13,gheorghiu16,rajpoot17}.
Such manifolds are determined by respective families of metrics $\mathbf{g}%
(\tau )=\mathbf{g}(\tau ,u)$ and N--connections $\mathbf{N}(\tau )=\mathbf{N}%
(\tau ,u)$ (we shall write only the parametric dependence if that will not
result in ambiguities) for which canonically corresponding d--connection
structures can be constructed $\mathbf{D}(\tau )=\mathbf{D}(\tau ,u).$ We
also suppose that on $\mathbf{V}(\tau )$ there are defined corresponding
families of Lagrange densities $\ ^{g}\mathcal{L}(\tau),$ for gravitational
fields in a MGT or GR, and $\ ^{tot}\mathcal{L}(\tau), $ as total
Lagrangians for effective and matter fields (\ref{lagrs}). For a double 2+2
and 3+1 splitting, we can consider local coordinates labeled as $u^{\alpha
}=(x^{i},y^{a})=(x^{\grave{\imath}},u^{4}=t)$ for $%
i,j,k,...=1,2;a,b,c,...=3,4;$ and $\grave{\imath},\grave{j},\grave{k}=1,2,3$%
. The nonholonomic distributions for N-connections can be parameterized
always in such forms that any open region $U\subset $ $\mathbf{V}$ is
covered by a family of 3-d spacelike hypersurfaces $\Xi _{t}$ parameterized
by a time like parameter $t.$

\subsubsection{Generalized Perelman functionals for entropic geometric flows
and MGTs}

For this class of theories, we postulate the modified Perelman's functionals
in the form
\begin{eqnarray}
\mathcal{F}(\tau ) &=&\int_{t_{1}}^{t_{2}}\int_{\Xi _{t}}e^{-f}\sqrt{|%
\mathbf{g}|}d^{4}u\left[ F(\ ^{s}R)+\ ^{tot}\mathcal{L}+|\mathbf{D}f|^{2}%
\right] \mbox{  and }  \label{fperelm4matter} \\
\mathcal{W}(\tau ) &=&\int_{t_{1}}^{t_{2}}\int_{\Xi _{t}}\left( 4\pi \tau
\right) ^{-3}e^{-f}\sqrt{|\mathbf{g}|}d^{4}u\left[ \tau \left( F(\ ^{s}R)+\
^{tot}\mathcal{L}+|h\mathbf{D}f|+|v\mathbf{D}f|\right) ^{2}+f-8\right] .
\label{wfperelm4matt}
\end{eqnarray}%
The condition $\int_{t_{1}}^{t_{2}}\int_{\Xi _{t}}\left( 4\pi \tau \right)
^{-3}e^{-f}\sqrt{|\mathbf{g}|}d^{4}u=1$ is imposed on the normalizing
function $f(\tau ,u).$ For topological considerations, such a normalisation
is not important. Nevertheless, it imposes certain nonholonomic constraints
on geometric objects which do not allow to solve derived geometric flow
evolution equations in explicit form. For applications to entropic gravity
and associated thermodynamic models, we can consider $\ f$ as an
undetermined scalar function which can be related to possible conformal
transforms or re-parameterizations. In result, we can prove certain general
decoupling and integration properties of corresponding systems of nonlinear
PDEs. Fixing a class of solutions, we can chose such integration functions
and constants which reproduce/ predict certain experimental and/or
observational data. Corresponding values of $\ f$ depend on systems of
reference and coordinates.

Let us explain and motivate the difference of (\ref{fperelm4matter}) and (%
\ref{wfperelm4matt}), introduced in the first partner work \cite{partner1},
from the original Grisha Perelman F- and W-functionals \cite{perelman1}
postulated for the Ricci flows of 3-d Riemannian metrics, see details in
monographs \cite{monogrrf1,monogrrf2,monogrrf3}. In this work, we study
geometric entropic flows of canonical geometric data $(\mathbf{g}(\tau ),%
\mathbf{N}(\tau ),\mathbf{D}(\tau ))$ for nonholonomic Lorentz manifolds and
various generalizations for MGTs following the program elaborated in \cite%
{bubuianu18,vacaru09,ruchin13,gheorghiu16,rajpoot17}, where possible
connections to emergent gravity were not analyzed. In formulas (\ref%
{fperelm4matter}) and (\ref{wfperelm4matt}), we consider the gravitational
Lagrangian $\ ^{g}\mathcal{L}=F(\ ^{s}R)$ as a functional of the scalar
curvature for $\mathbf{D}$, or $\ ^{g}\mathcal{L}=R[\nabla ]$ for
considering as particular cases models of geometric evolution of exact
solutions in GR. The key difference from previous works is that in such
relativistic functionals the term $\ ^{tot}\mathcal{L}$ is introduced, which
is responsible for geometric evolution of configurations with elasticity and
quasiperiodicity. Those functionals can be generalized on a temperature like
parameter $\tau $ and used as certain alternative geometric functionals, for
instance, for W-entropy. Nevertheless, only nonholonomic elastic
quasiperiodic functionals of type (\ref{fperelm4matter}) and (\ref%
{wfperelm4matt}) result for self-similar configurations (see next
subsections) in entropic gravity equations of E. Verlinde type \cite%
{verlinde10,verlinde16,hossenfelder17,partner1} and/or with quasiperiodic
structure \cite{vacaru18tc,bubuianu18,amaral17}.

In this and partner \cite{partner1} papers, we work with generalized
geometric flow and entropy functionals determined by $F(\ ^{s}R)+\ ^{tot}%
\mathcal{L}$ and $\mathbf{D,}$ respectively, instead of the Riemannian
values $R$ and $\nabla $ used in the former mathematical works. In our
nonholonomic approaches, above F- and W--functionals characterize
relativistic thermodynamic models with analogous nonlinear hydrodynamic
flows of families of entropic values, metrics and generalized connections,
encoding interactions of gravitational and matter fields as it is motivated
in \cite{ruchin13,gheorghiu16,rajpoot17}. In general, it is possible to work
with any class of normalizing functions $f(\tau ,u)$ which can be redefined
in order to include geometric and matter Lagrange terms and certain constant
values and parameters. In many cases, such a function is chosen in a
non--explicit form. This allows us to study non--normalized geometric flows
but with nonholonomic constraints. For such conditions, there found various
general decoupling and integration properties of respective physically
important systems of nonlinear PDEs. In result, generic off-diagonal
solutions can be constructed in explicit form as in \cite%
{partner1,bubuianu18,vacaru09,ruchin13,gheorghiu16,rajpoot17}, but with
entropic and quasiperiodic modifications. The existence of such solutions
validates our nonholonomic geometric flow entropic approach, involving
metrics with pseudo-Euclidean signature even analogs of the Poincar\'{e}%
--Thurston conjecture have not been formulated and proven for the Lorentzian
spacetimes. Nevertheless, explicit constructions of exact solutions with
elastic and quasiperiodic gravitational metrics and effective matter
sources, which will be provided in sections \ref{s4}-\ref{s6}, support E.
Velinde conjecture on entropic gravity which in our works is proven for
modified Poincar\'{e} functionals.

\subsubsection{Nonholonomic 3-d space like hypersurface F- and W-functionals}

We can redefine and compute relativistic entropies (\ref{fperelm4matter})
and (\ref{wfperelm4matt}) for any 3+1 splitting with 3-d closed hypersurface
fibrations $\widehat{\Xi }_{t}$ as we described above in section \ref{ss31sp}%
.

Let us denote by$\ _{\shortmid }\mathbf{D}=\mathbf{D}_{\mid \widehat{\Xi }%
_{t}}$ the canonical d--connection $\mathbf{D}$ defined on a 3-d
hypersurface $\widehat{\Xi }_{t},$ when all values depend on a temperature
like parameter $\tau (\tau ^{\prime })$ with possible scale re-definitions
for another parameter $\tau ^{\prime }$ etc. We define also $\ _{\shortmid
}^{s}R:=$ $\ ^{s}R_{\mid \widehat{\Xi }_{t}}.$ Using $q_{\grave{\imath}%
}(\tau )=[q_{i}(\tau ),q_{3}(\tau )]$ in a family of d-metrics (\ref%
{offdublespl}), the Perelman's functionals parameterized in N--adapted form
are constructed in the form:
\begin{eqnarray}
\ _{\shortmid }\mathcal{F} &=&\int_{\widehat{\Xi }_{t}}e^{-\ _{\shortmid }f}%
\sqrt{|q_{\grave{\imath}\grave{j}}|}d\grave{x}^{3}\left[ (\ _{\shortmid }F(\
_{\shortmid }^{s}R)+\ _{\shortmid }^{tot}\mathcal{L+}|\ _{\shortmid }\mathbf{%
D}\ _{\shortmid }f|^{2})\right] ,\mbox{ and }  \label{perelm3f} \\
\ _{\shortmid }\mathcal{W} &=&\int_{\widehat{\Xi }_{t}}\ _{\shortmid }\mu
\sqrt{|q_{\grave{\imath}\grave{j}}|}d\grave{x}^{3}\left[ \tau \left( (\
_{\shortmid }F(\ _{\shortmid }^{s}R)+\ _{\shortmid }^{tot}\mathcal{L}+|\ \
_{\shortmid }^{h}\mathbf{D}f|+|\ \ _{\shortmid }^{v}\mathbf{D}\ _{\shortmid
}f|\right) ^{2}+\ _{\shortmid }f-6\right] .  \label{perelm3w}
\end{eqnarray}%
These functionals are derived respectively from the previous 4-d elastic
ones when the values $\ _{\shortmid }F(\ _{\shortmid }^{s}R)$ and$\
_{\shortmid }^{tot}\mathcal{L}$ are computed as projections on a 3-d
hypersurface for a redefined normalization function $_{\shortmid }f.$ Using
frame/coordinate transform and re-definition of the temperature like
parameter, we can always chose a necessary type scaling function $\
_{\shortmid }f$ which satisfies normalization conditions $\int_{\widehat{\Xi
}_{t}}\ _{\shortmid }\mu \sqrt{|q_{\grave{\imath}\grave{j}}|}d\grave{x}%
^{3}=1 $ for $\ _{\shortmid }\mu =\left( 4\pi \tau \right) ^{-3}e^{-\
_{\shortmid }f}.$

The functionals (\ref{perelm3f}) and (\ref{perelm3w})\ transform into
standard Perelman functionals \cite{perelman1} for 3-d Riemannian metrics on
$\widehat{\Xi }_{t}$ if $\ _{\shortmid }\mathbf{D}\rightarrow \ _{\shortmid
}\nabla ,\ _{\shortmid }F(\ _{\shortmid }^{s}R)=\ _{\shortmid }^{s}R$ and $\
_{\shortmid }^{tot}\mathcal{L}=0.$ In order to describe possible
contributions on 3-d hypersurfaces of spacetime elasticity and quasiperiodic
structure in entropic gravity, it is necessary to analyze physical effects
of such nonholonomic deformations.

\subsection{Geometric flow equations for modified gravitational \& matter
fields}

Applying a variational procedure for a corresponding F-functional for
geometric flows of 3-d Riemannian metrics, G. Perelman \cite{perelman1}
provided a proof for R. Hamilton's equations \cite{hamilt1,hamilt2,hamilt3}.
For self-consistent configurations with a fixed flow parameter $\tau _{0},$
one obtains 3-d Ricci soliton equations which are equivalent to the vacuum
Einstein equations for $\nabla $ with an effective cosmological constant. In
similar forms to rigorous mathematical proofs in \cite%
{perelman1,monogrrf1,monogrrf2,monogrrf3} but elaborating on N-adapted
variational procedures, for instance, for the functional $\mathcal{F}(\tau )$
(\ref{fperelm4matter}) with a canonical $\mathbf{D}$ used instead of $\nabla
$ (see details in \cite{ruchin13,gheorghiu16,rajpoot17,partner1}), we obtain
a system of nonlinear PDEs generalizing the R. Hamilton equations for
entropic and quasiperiodic geometric flow evolution determined by canonical
data $(\mathbf{g=\{g}_{\mu \nu }=[g_{ij},g_{ab}]\},\mathbf{N=\{}N_{i}^{a}\},%
\mathbf{D},\ ^{tot}\mathcal{L}),$
\begin{eqnarray}
\partial _{\tau }g_{ij} &=&-2(\mathbf{R}_{ij}-\ ^{tot}\Upsilon _{ij});\
\label{ricciflowr2} \\
\partial _{\tau }g_{ab} &=&-2(\mathbf{R}_{ab}-\ ^{tot}\Upsilon _{ab});
\notag \\
\mathbf{R}_{ia} &=&\mathbf{R}_{ai}=0;\mathbf{R}_{ij}=\mathbf{R}_{ji};\mathbf{%
R}_{ab}=\mathbf{R}_{ba};  \notag \\
\partial _{\tau }f &=&-\widehat{\square }f+\left\vert \mathbf{D}f\right\vert
^{2}-\ ^{s}R+\ ^{tot}\Upsilon _{\alpha }^{\alpha }).  \notag
\end{eqnarray}%
In these formulas, $\widehat{\square }(\tau )=\mathbf{D}^{\alpha }(\tau )%
\mathbf{D}_{\alpha }(\tau )$ and $\ ^{tot}\Upsilon _{\alpha \beta }(\tau )$
is chosen for geometric flows of (effective) sources of entropic gravity (%
\ref{totsourc}) (if we fix any $\tau =\tau _{0}).$ We note that the
dependence on a flow parameter $\tau $ for such (effective) matter sources
is determined by certain evolutions of $\mathbf{g}(\tau )$ and $\mathbf{D}%
(\tau ).$ In such theories, we do not consider nonholonomic deformations and
evolution of classical matter fields. For instance, we do not consider
geometric flow evolution equations for the electromagnetic potential $%
\mathbf{A}_{\alpha }(\tau )$ with evolution terms of type $\partial _{\tau }%
\mathbf{A}_{\alpha }$ even such theories were studied in our previous works
\cite%
{vacaru06,vacaru07,vacaru09,vacaru10,vacaru13,ruchin13,gheorghiu16,rajpoot17,bubuianu18a}%
, see references therein.

The conditions $\mathbf{R}_{ia}=0$ and $\mathbf{R}_{ai}=0$ for the Ricci
tensor $Ric[\mathbf{D}]=\{\mathbf{R}_{\alpha \beta }=[\mathbf{R}_{ij},%
\mathbf{R}_{ia},\mathbf{R}_{ai},\mathbf{R}_{ab}]\}$ are necessary if we want
to keep the metric $\mathbf{g}(\tau )$ to be symmetric under nonholonomic
Ricci flow evolution determined by (\ref{ricciflowr2}). Geometric flow
evolution and nonholonomic gravity of theories with nonsymmetric metrics
were studied in \cite{v09}, see references therein. In principle, we can
work with any type normalization function $f$ which allows a general
decoupling and integration of such systems of nonlinear PDEs. Such a
normalization depends on frame and coordinate transforms and may encode
(effective) cosmological constants, matter sources etc. We note that similar
variational and/or geometric methods allows to derive from $\mathcal{W}(\tau
)$ (\ref{wfperelm4matt}) certain types nonlinear evolution equations which
are equivalent to (\ref{ricciflowr2}). It is more difficult to solve
explicitly such PDEs but a W-functional allows to elaborate directly on
certain classes of thermodynamic models, see section \ref{s5}.

\subsection{Entropic gravity and gravitational field equations as Ricci
solitons}

\label{ssentrg}

For self-similar point $\tau =\tau _{0}$ configurations when $\partial
_{\tau }\mathbf{g}_{\mu \nu }=0,$ with a corresponding choice of the
normalizing geometric flow function $f,$ the equations (\ref{ricciflowr2})
transform into relativistic nonholonomic Ricci soliton equations%
\begin{equation}
\mathbf{R}_{ij} =\ ^{tot}\Upsilon _{ij},\ \mathbf{R}_{ab} =\ ^{tot}\Upsilon
_{ab}, \ \mathbf{R}_{ia} = \mathbf{R}_{ai}=0  \label{entrsolit}
\end{equation}%
which are equivalent to (modified) Einstein equations in (MGT) GR for
corresponding definitions of $\ ^{tot}\Upsilon _{\alpha \beta }.$ A class of
MGTs and GR can be formulated as geometric theories of entropic elastic
origin which is similar to the idea of emergent gravity put forward by E.
Verlinde \cite{verlinde10,verlinde16}, i.e. in the form (\ref{deinst}) with
(effective) entropic and quasiperiodic source $\ ^{tot}\Upsilon
_{\alpha\beta }$ (\ref{totsourc}).

We conclude that an emergent gravity model in the E. Verlinde sense \cite%
{verlinde10,verlinde16,hossenfelder17}, can be constructed for Lagrange
distributions (\ref{lagrs}) and respective sources (\ref{totsourc})
introduced as generating data for the nonholonomic Hamilton equations (\ref%
{ricciflowr2}) and respective relativistic Ricci solitons. Such geometric
flow evolution theories and their spacetime elastic, quasiperiodic and
thermodynamic properties are determined by the generalized $\mathcal{W}$%
--entropy (\ref{wfperelm4matt}).

\section{Decoupling and integrability of entropic flow equations}

\label{s4}In this section, we prove that the system of nonlinear PDEs (\ref%
{ricciflowr2}) describing spacetime elastic and quasiperiodic flows and
entropic gravity theories can be formally integrated in very general forms
for generic off-diagonal metrics and canonical d-connections (in particular,
for LC-configurations). The coefficients of geometric objects for such
solutions depend on all spacetime coordinates via generating and integration
functions and (effective) matter sources. The anholonomic frame deformation
method, AFDM, for constructing exact solutions in MGTs and GR is developed
for generating new classes of solutions encoding entropic quasiperiodic
modifications in $\mathbf{g}(\tau )$ (\ref{dm}), $\mathbf{D}(\tau )$ (\ref%
{lcconcdcon}), and $\ ^{tot}\Upsilon _{\alpha \beta }(\tau )$ (\ref{totsourc}%
). For similar details and mathematical proofs, we refer readers to our
previous works \cite{v05,v09,v09a,gheorghiu14,bubuianu17,v16,
bubuianu18,vacaru18tc}, on exact solutions in MGTs, and \cite%
{vacaru06,vacaru07,vacaru09,vacaru10,vacaru13,ruchin13,gheorghiu16,rajpoot17,bubuianu18a}%
, for solutions with nonholonomic Ricci flows, and citations therein.

\subsection{Geometric flows with parametric modified Einstein equations}

Introducing effective sources, entropic geometric flow equations can written
as modified Einstein equations with dependence on a temperature like
parameter $\tau$. We show that such systems of nonlinear PDEs can be
decoupled in general forms.

\subsubsection{Entropic quasiperiodic flow modifications of gravitational
field equations}

Using nonholonomic frame transforms and tetradic (vierbein) fields, we
introduce effective sources which in N--adapted form are parameterized
\begin{equation}
\ ^{eff}\yen _{\mu \nu }(\tau )=\mathbf{e}_{\ \mu }^{\mu ^{\prime }}(\tau )%
\mathbf{e}_{\nu }^{\ \nu ^{\prime }}(\tau )[~\ ^{tot}\mathbf{\Upsilon }_{\mu
^{\prime }\nu ^{\prime }}(\tau )+\frac{1}{2}~\partial _{\tau }\mathbf{g}%
_{\mu ^{\prime }\nu ^{\prime }}(\tau )]=[~\ _{h}\yen (\tau ,{x}^{k})\delta
_{j}^{i},\yen (\tau ,x^{k},y^{c})\delta _{b}^{a}].  \label{effsourc}
\end{equation}%
Such families of vielbein transforms $\mathbf{e}_{\ \mu ^{\prime }}^{\mu
}(\tau )=\mathbf{e}_{\ \mu ^{\prime }}^{\mu }(\tau ,u^{\gamma })$ and their
dual $\mathbf{e}_{\nu }^{\ \nu ^{\prime }}(\tau ,u^{\gamma })$, when $%
\mathbf{e}_{\ }^{\mu }=\mathbf{e}_{\ \mu ^{\prime }}^{\mu }du^{\mu ^{\prime
}}$ can be chosen for any frame/coordinate transforms of a N-splitting
structure (\ref{whitney}). In result, the system of nonholonomic entropic R.
Hamilton equations (\ref{ricciflowr2}) can be written in the form (\ref%
{deinst}) but with geometric objects depending additionally on a temperature
like parameter $\tau $ and for effective source (\ref{effsourc}),%
\begin{equation}
\mathbf{R}_{\alpha \beta }(\tau )=\ ^{eff}\yen _{\alpha \beta }(\tau ).
\label{entropfloweq}
\end{equation}%
We note that such geometric evolution equations are for an undetermined
normalization function $f(\tau )=f(\tau ,(\tau ,u^{\gamma })$ which can be
defined explicitly for respective classes of exact or parametric solutions.
For self-similar point $\tau =\tau _{0}$ configurations with $\partial
_{\tau }\mathbf{g}_{\mu \nu }(\tau _{0})=0,$ this system of nonlinear PDEs
transforms into the nonholonomic entropic Ricci soliton equations (\ref%
{entrsolit}).

\subsubsection{Effective entropic sources for stationary and/or cosmological
configurations}

The values $\ _{h}\yen (\tau ,{x})$ and $\yen (\tau ,x,y)$\ in (\ref%
{effsourc}) can be considered as generating data for (effective) matter
sources. Prescribing such data, we impose certain nonholonomic frame
constraints on geometric evolution and self-similar configurations of
entropic and quasiperiodic structures. This type of $\yen $--generating
functions allows formal integrations of the system (\ref{entropfloweq}) in
certain general forms.

Using frame transforms, the $\tau $-evolution of d--metric $\mathbf{g}(\tau
) $ (\ref{dm}) can be parameterized for respective spherical symmetric
coordinates $u^{\alpha }=({r,\theta },y^{3}=\varphi ,t)$ or some
cosmological coordinates $(x^{k},y^{4}=t)$,
\begin{eqnarray}
g_{i}(\tau ) &=&e^{\psi {(\tau ,r,\theta )}},\,\,\,\,g_{a}(\tau )=\omega ({%
\tau ,r,\theta },y^{b})h_{a}({\tau ,r,\theta },\varphi ),  \label{statf} \\
\ N_{i}^{3}(\tau ) &=&w_{i}({\tau ,r,\theta },\varphi
),\,\,\,\,N_{i}^{4}(\tau )=n_{i}({\tau ,r,\theta },\varphi ),\mbox{ for }%
\omega =1,\mbox{stationary configurations };  \notag \\
g_{i}(\tau ) &=&e^{\psi {(\tau ,x^{k})}},\,\,\,\,g_{a}(\tau )=\omega ({\tau ,%
}x^{k},y^{b})\overline{h}_{a}({\tau ,}x^{k},t),\   \label{cosmf} \\
N_{i}^{3}(\tau ) &=&\overline{n}_{i}({\tau ,}x^{k},t),\,\,\,\,N_{i}^{4}(\tau
)=\overline{w}_{i}({\tau ,}x^{k},t),\mbox{ for }\omega =1,%
\mbox{
cosmological configurations }  \notag
\end{eqnarray}%
The AFDM results in more simple and explicit (still very general classes) of
solutions if we work with nonholonomic configurations possessing at least
one Killing symmetry, for instance, on $\partial _{4}=\partial _{t}$ for
stationary solution or on $\partial _{3}=\partial _{\varphi },$ locally
anisotropic solutions.\footnote{%
In principle, we can construct for (\ref{entropfloweq}) certain classes of
exact and parametric off-diagonal solutions generically depending on all
spacetime coordinates $(x^{k},y^{a})$ but that would result in hundreds of
pages with a cumbersome formulas for respective geometric techniques, see
\cite{v05,v09,v09a,gheorghiu14,bubuianu17} and references therein.}

We shall use brief notations of partial derivatives $\partial
_{\alpha}q=\partial q/\partial u^{\alpha }$ when a function $q(x^{k},y^{a}),$
\begin{eqnarray*}
\partial _{1}q &=&q^{\bullet }=\partial q/\partial x^{1},\partial
_{2}q=q^{\prime }=\partial q/\partial x^{2},\partial _{3}q=\partial
q/\partial y^{3}=\partial q/\partial \varphi =q^{\diamond },\partial
_{4}q=\partial q/\partial t=\partial _{t}q=q^{\ast }, \\
\partial _{33}^{2} &=&\partial ^{2}q/\partial \varphi ^{2}=\partial
_{\varphi \varphi }^{2}q=q^{\diamond \diamond },\partial _{44}^{2}=\partial
^{2}q/\partial t^{2}=\partial _{tt}^{2}q=q^{\ast \ast }.
\end{eqnarray*}%
For respective Killing symmetries, the effective sources $\yen (\tau ,x,y)$
in (\ref{effsourc}) can be parameterized
\begin{equation}
\ ^{eff}\yen _{\ \nu }^{\mu }(\tau )=\left\{
\begin{array}{cc}
\lbrack ~\ _{h}\yen (\tau ,{r,\theta })\delta _{j}^{i},\yen (\tau ,{r,\theta
},\varphi )\delta _{b}^{a}], & \mbox{ stationary configurations }; \\
\lbrack ~\ _{h}\overline{\yen }(\tau ,x^{i})\delta _{j}^{i},\overline{\yen }%
(\tau ,x^{i},t)\delta _{b}^{a}], & \mbox{ cosmological configurations }.%
\end{array}%
\right.  \label{dsourcparam}
\end{equation}%
Considering as typical examples two types of a Killing space symmetry or
time like Killing symmetry for effective generating sources, we shall
construct and study properties of two general classes of exact solutions
(the first one will be for stationary configurations which may contain BH
solutions and the second one will be for cosmological type solutions).

\subsection{Nontrivial Ricci d-tensors and decoupling of entropic flow
equations}

In this subsection, we outline the key steps for proofs of general
decoupling and integrability of (modified) Einstein equations with effective
sources (\ref{dsourcparam}).

\subsubsection{Off--diagonal metric ansatz, (non) holonomic variables, and
ODEs and PDEs}

Let us summarize in Table 1 below the data on nonholonomic 3+1 and 2+2
variables and corresponding ansatz which allows to transform geometric and
entropic flow equations and, a nonholonomic Ricci solitons, gravitational
field equations in entropic MGTs and GR into respective systems of nonlinear
ordinary differential equations, ODEs, and partial differential equations,
PDEs. All formulas will be proven in next subsections. We model a
nonholonomic deformation with $\eta $-polarization functions, $\mathbf{%
\mathring{g}\rightarrow g}(\tau ),$ of a 'prime' metric, $\mathbf{\mathring{g%
}}$, into a family 'target' d-metrics $\mathbf{g}(\tau )$ (\ref{dm}), if
{\small
\begin{equation}
\mathbf{g}(\tau )=\eta _{i}(\tau ,x^{k})\mathring{g}_{i}dx^{i}\otimes
dx^{i}+\eta _{a}(\tau ,x^{k},y^{b})\mathring{h}_{a}\mathbf{e}^{a}[\eta
]\otimes \mathbf{e}^{a}[\eta ],  \label{dme}
\end{equation}%
} where the target N-elongated basis is determined by $N_{i}^{a}(\tau
,u)=\eta _{i}^{a}(\tau ,x^{k},y^{b})\mathring{N}_{i}^{a}(\tau ,x^{k},y^{b})$
in the form\footnote{%
we do not consider summation on repeating indices if they are not written as
contraction of "up-low" ones} $\mathbf{e}^{\alpha }[\eta ]=(dx^{i},\mathbf{e}%
^{a}=dy^{a}+\eta _{i}^{a}\mathring{N}_{i}^{a}dx^{i})$. The values $\eta
_{i}(\tau )=\eta _{i}(\tau ,x^{k}),\eta _{a}(\tau )=\eta _{a}(\tau
,x^{k},y^{b})$ and $\eta _{i}^{a}(\tau )=\eta _{i}^{a}(\tau ,x^{k},y^{b})$
are called respectively geometric/entropic flow or gravitational
polarization functions, or $\eta $-polarizations. Any $\mathbf{g}(\tau )$ is
subjected to the condition that it defines a solution of modified Einstein
equations resulting in entropic quasiperiodic geometric flows and/or via
nonholonomic deformations. A general prime metric in a coordinate
parametrization is of type $\mathbf{\mathring{g}}=\mathring{g}_{\alpha \beta
}(x^{i},y^{a})du^{\alpha }\otimes du^{\beta },$ which can be also
represented equivalently in N-adapted form
\begin{eqnarray}
\mathbf{\mathring{g}} &=&\mathring{g}_{\alpha }(u)\mathbf{\mathring{e}}%
^{\alpha }\otimes \mathbf{\mathring{e}}^{\beta }=\mathring{g}%
_{i}(x)dx^{i}\otimes dx^{i}+\mathring{g}_{a}(x,y)\mathbf{\mathring{e}}%
^{a}\otimes \mathbf{\mathring{e}}^{a},  \label{primedm} \\
&&\mbox{ for }\mathbf{\mathring{e}}^{\alpha }=(dx^{i},\mathbf{e}^{a}=dy^{a}+%
\mathring{N}_{i}^{a}(u)dx^{i}),\mbox{ and }\mathbf{\mathring{e}}_{\alpha }=(%
\mathbf{\mathring{e}}_{i}=\partial /\partial y^{a}-\mathring{N}%
_{i}^{b}(u)\partial /\partial y^{b},\ {e}_{a}=\partial /\partial y^{a}).
\notag
\end{eqnarray}%
Such a d-metric can be, or not, a solution of some gravitational field
equations in a MGT or GR but it nonholonomic deformations to a target metric
(\ref{dme}) are subjected to the condition to define an exact or parametric
solutions of certain entropic flow evolutions equations.

%%%%%%%%%%%
% Table 1
%%%%%%%%%%%

%\vskip5pt
%\begin{table*}[h]
{\scriptsize
\begin{eqnarray*}
&&%
\begin{tabular}{l}
\hline\hline
\begin{tabular}{lll}
& {\ \textsf{Table 1:\ Entropic quasiperiodic flow modified Einstein eqs as
systems of nonlinear PDEs}} &  \\
& and the Anholonomic Frame Deformation Method, \textbf{AFDM}, &  \\
& \textit{for constructing generic off-diagonal exact, parametric, and
physically important solutions} &
\end{tabular}%
\end{tabular}
\\
&&{%
\begin{tabular}{lll}
\hline
diagonal ansatz: PDEs $\rightarrow $ \textbf{ODE}s &  & AFDM: \textbf{PDE}s
\textbf{with decoupling; \ generating functions} \\
radial coordinates $u^{\alpha }=(r,\theta ,\varphi ,t)$ & $u=(x,y):$ &
\mbox{  2+2
splitting, } $u^{\alpha }=(x^{1},x^{2},y^{3},y^{4}=t);$%
\mbox{  flow
parameter  }$\tau $ \\
LC-connection $\mathring{\nabla}$ & [connections] & $%
\begin{array}{c}
\mathbf{N}:T\mathbf{V}=hT\mathbf{V}\oplus vT\mathbf{V,}\mbox{ locally }%
\mathbf{N}=\{N_{i}^{a}(x,y)\} \\
\mbox{ canonical connection distortion }\mathbf{D}=\nabla +\mathbf{Z}%
\end{array}%
$ \\
$%
\begin{array}{c}
\mbox{ diagonal ansatz  }g_{\alpha \beta }(u) \\
=\left(
\begin{array}{cccc}
\mathring{g}_{1} &  &  &  \\
& \mathring{g}_{2} &  &  \\
&  & \mathring{g}_{3} &  \\
&  &  & \mathring{g}_{4}%
\end{array}%
\right)%
\end{array}%
$ & $\mathbf{\mathring{g}}\Leftrightarrow \mathbf{g}(\tau )$ & $%
\begin{array}{c}
g_{\alpha \beta }(\tau )=%
\begin{array}{c}
g_{\alpha \beta }(\tau ,x^{i},y^{a})\mbox{ general frames / coordinates} \\
\left[
\begin{array}{cc}
g_{ij}+N_{i}^{a}N_{j}^{b}h_{ab} & N_{i}^{b}h_{cb} \\
N_{j}^{a}h_{ab} & h_{ac}%
\end{array}%
\right] ,\mbox{ 2 x 2 blocks }%
\end{array}
\\
\mathbf{g}_{\alpha \beta }(\tau )=[g_{ij}(\tau ),h_{ab}(\tau )], \\
\mathbf{g}(\tau )=\mathbf{g}_{i}(\tau ,x^{k})dx^{i}\otimes dx^{i}+\mathbf{g}%
_{a}(\tau ,x^{k},y^{b})\mathbf{e}^{a}\otimes \mathbf{e}^{b}%
\end{array}%
$ \\
$\mathring{g}_{\alpha \beta }=\left\{
\begin{array}{cc}
\mathring{g}_{\alpha }(r) & \mbox{ for BHs} \\
\mathring{g}_{\alpha }(t) & \mbox{ for FLRW }%
\end{array}%
\right. $ & [coord.frames] & $g_{\alpha \beta }(\tau )=\left\{
\begin{array}{cc}
g_{\alpha \beta }(\tau ,r,\theta ,y^{3}=\varphi ) &
\mbox{ stationary
configurations} \\
g_{\alpha \beta }(\tau ,r,\theta ,y^{4}=t) & \mbox{ cosm. configurations}%
\end{array}%
\right. $ \\
&  &  \\
$%
\begin{array}{c}
\mbox{coord.tranfsorms }e_{\alpha }=e_{\ \alpha }^{\alpha ^{\prime
}}\partial _{\alpha ^{\prime }}, \\
e^{\beta }=e_{\beta ^{\prime }}^{\ \beta }du^{\beta ^{\prime }},\mathring{g}%
_{\alpha \beta }=\mathring{g}_{\alpha ^{\prime }\beta ^{\prime }}e_{\ \alpha
}^{\alpha ^{\prime }}e_{\ \beta }^{\beta ^{\prime }} \\
\begin{array}{c}
\mathbf{\mathring{g}}_{\alpha }(x^{k},y^{a})\rightarrow \mathring{g}_{\alpha
}(r),\mbox{ or }\mathring{g}_{\alpha }(t), \\
\mathring{N}_{i}^{a}(x^{k},y^{a})\rightarrow 0.%
\end{array}%
\end{array}%
$ & [N-adapt. fr.] & $\left\{
\begin{array}{cc}
\begin{array}{c}
\mathbf{g}_{i}(\tau ,r,\theta ),\mathbf{g}_{a}(\tau ,r,\theta ,\varphi ), \\
\mbox{ or }\mathbf{g}_{i}(\tau ,r,\theta ),\mathbf{g}_{a}(\tau ,r,\theta ,t),%
\end{array}
& \mbox{ d-metrics } \\
\begin{array}{c}
N_{i}^{3}(\tau )=w_{i}(\tau ,r,\theta ,\varphi ),N_{i}^{4}=n_{i}(\tau
,r,\theta ,\varphi ), \\
\mbox{ or }N_{i}^{3}(\tau )=n_{i}(\tau ,r,\theta ,t),N_{i}^{4}=w_{i}(\tau
,r,\theta ,t),%
\end{array}
&
\end{array}%
\right. $ \\
$\mathring{\nabla},$ $Ric=\{\mathring{R}_{\ \beta \gamma }\}$ & Ricci tensors
& $\mathbf{D},\ \mathcal{R}ic=\{\mathbf{R}_{\ \beta \gamma }\}$ \\
$~^{m}\mathcal{L[\mathbf{\phi }]\rightarrow }\ ^{m}\mathbf{T}_{\alpha \beta }%
\mathcal{[\mathbf{\phi }]}$ & sources & $%
\begin{array}{cc}
\mathbf{\Upsilon }_{\ \nu }^{\mu }(\tau )=\mathbf{e}_{\ \mu ^{\prime }}^{\mu
}\mathbf{e}_{\nu }^{\ \nu ^{\prime }}\mathbf{\Upsilon }_{\ \nu ^{\prime
}}^{\mu ^{\prime }} &  \\
=diag[~\ _{h}\yen (\tau ,x^{i})\delta _{j}^{i},\yen (\tau ,x^{i},\varphi
)\delta _{b}^{a}], & \mbox{stationary conf.} \\
=diag[~\ _{h}\overline{\yen }(\tau ,x^{i})\delta _{j}^{i},\overline{\yen }%
(\tau ,x^{i},t)\delta _{b}^{a}], & \mbox{ cosmol. conf.}%
\end{array}%
$ \\
trivial equations for $\mathring{\nabla}$-torsion & LC-conditions & $\mathbf{%
D}_{\mid \widehat{\mathcal{T}}\rightarrow 0}=\mathbf{\nabla }
\mbox{
extracting new classes of solutions in GR}$ \\ \hline\hline
\end{tabular}%
}
\end{eqnarray*}%
}%\caption{Off-diagonal ansatz, connections and sources for the AFDM}
%\label{tb1par}
%\end{table*}

In our works, we are interested usually in two physically important cases
when $\mathbf{\mathring{g}}$ (\ref{primedm}) defines a BH solution (for
instance, a vacuum Kerr, or Schwarzschild, Kerr-(anti) de Sitter metric), or
a Friedman--Lema\^{\i}tre--Robertson--Walker (FLRW) type metric, or any
Bianchi anisotropic metrics. For diagonalizable prime metrics (the
off-diagonal structure of the Kerr metric is determined by rotation frames
and coordinates), we can always find a coordinate system when $\mathring{N}%
_{i}^{b}=0.$ To avoid noholonomic deformations with singular coordinates is
convenient to construct exact solutions with nontrivial functions $\eta
_{\alpha }=(\eta _{i},\eta _{a}),\eta _{i}^{a},$ and nonzero coefficients $%
\mathring{N}_{i}^{b}(u).$ We have to consider necessary type frame/
coordinate transforms. For a d-metric (\ref{dme}), we can analyze the
conditions of existence and geometric/ physical properties of some target
and/or prime solutions, for instance, when $\eta _{\alpha }\rightarrow 1$
and $N_{i}^{a}\rightarrow \mathring{N}_{i}^{a}.$ The values $\eta _{\alpha
}=1$ and/or $\mathring{N}_{i}^{a}=0$ can be imposed as some special
nonholonomic constraints.\footnote{\label{fnsmallp}We can consider flow
evolution of a physical importan target metric $\mathbf{g}$ (\ref{dme}) with
generic off-diagonal terms as an "almost" BH, or FLRW cosmological, like
metric. Such parametric solutions are constructed for small nonholonomic
deformations on some constant parameters $\eta _{\alpha }=(\eta _{i},\eta
_{a}),\eta _{i}^{a},$ for $0\leq \varepsilon _{\alpha },\varepsilon
_{i}^{b}\ll 1,$ when $\eta _{i}\simeq \check{\eta}_{i}(\tau
,x^{k})[1+\varepsilon _{i}\chi _{i}(\tau ,x^{k})]\simeq 1+\varepsilon
_{i}\chi _{i}(\tau ,x^{k}),\eta _{a}\simeq \check{\eta}_{a}(\tau
,x^{k},y^{b})[1+\varepsilon _{a}\chi _{a}(\tau ,x^{k},y^{b})]\simeq
1+\varepsilon _{a}\chi _{a}(\tau ,x^{k},y^{b})$, and $\eta _{i}^{a}\simeq
\check{\eta}_{i}^{a}(\tau ,x^{k},y^{b})[1+\varepsilon _{i}^{a}\chi
_{i}^{a}(\tau ,x^{k},y^{b})]\simeq 1+\varepsilon _{i}^{a}\chi _{i}^{a}(\tau
,x^{k},y^{b})$. Parametric $\varepsilon $-decompositions can be performed in
a self-consistent form by omitting quadratic and higher terms after a class
of solutions have been found for some evolution or nonholonomic deformation
data $(\eta _{\alpha },\eta _{i}^{a}).$ For certain subclasses of solutions,
we can consider that $\varepsilon _{i},\varepsilon _{a},\varepsilon
_{i}^{a}\sim \varepsilon $, when only one small parameter is considered for
all coefficients of nonholonomic deformations. We can work with mixed types
of solutions and model only small diagonal deformations $\varepsilon
_{i},\varepsilon _{a},\sim \varepsilon $ of metrics, for some general $\eta
_{i}^{a}.$ Alternatively, we consider nontrivial $\eta _{\alpha }$ but $%
\varepsilon _{i}^{a}$ $\sim \varepsilon .$} In brief, we shall denote
certain nonholonomic entropic deformations of a prime d-metrics into a
target one as $\mathbf{\mathring{g}}\rightarrow \mathbf{g}=[g_{\alpha }=\eta
_{\alpha }\mathring{g}_{\alpha },\ \eta _{i}^{a}\mathring{N}_{i}^{a}].$

Table 1 outlines the key steps for developing the AFDM to theories of
entropic quasiperiodic geometric flows. In this work, the formulas depend on
a flow temperature like parameter $\tau$ and the constructions are for
effective matter sources encoding entropic nonholonomic flows and
deformations.

\subsubsection{Cosmological Ricci d-tensors, LC--conditions, and nonlinear
symmetries}

\label{ssdeccosm}For locally anisotropic cosmological configurations, we can
consider geometric data when coefficients of the geometric objects do not
depend on a space like $y^{3}$ with respect to certain classes of N-adapted
frames. Using d-metric data (\ref{cosmf}) with $\omega =1$ and a source $%
\left[ \ _{h}\overline{\yen }(\tau ,x^{i}),\overline{\yen }(\,\tau ,x^{i},t)%
\right] $ \ (\ref{dsourcparam}), we write the entropic flow modified
Einstein equations (\ref{entropfloweq}) in the form\footnote{%
there are considered partial derivatives $\partial _{t}q=$ $\partial
_{4}q=q^{\ast }$ and $\partial _{i}q=(\partial _{1}q=q^{\bullet },\partial
_{2}q=q^{\prime });$ we use over-lined symbols in order to emphasize that
certain values are not stationary but depend on a time like coordinate $t$}:
{\small
\begin{eqnarray}
\mathbf{R}_{1}^{1}(\tau ) &=&\mathbf{R}_{2}^{2}(\tau )=-\ \ _{h}\overline{%
\yen }(\tau )\mbox{
i.e. }\frac{g_{1}^{\bullet }g_{2}^{\bullet }}{2g_{1}}+\frac{\left(
g_{2}^{\bullet }\right) ^{2}}{2g_{2}}-g_{2}^{\bullet \bullet }+\frac{%
g_{1}^{\prime }g_{2}^{\prime }}{2g_{2}}+\frac{(g_{1}^{\prime })^{2}}{2g_{1}}%
-g_{1}^{\prime \prime }=-2g_{1}g_{2}\ _{h}\overline{\yen };  \label{eq1b} \\
\mathbf{R}_{3}^{3}(\tau ) &=&\mathbf{R}_{4}^{4}(\tau )=-\overline{\yen }%
(\tau )\mbox{  i.e. }\frac{\left( \overline{h}_{3}^{\ast }\right) ^{2}}{2%
\overline{h}_{3}}+\frac{\overline{h}_{3}^{\ast }\overline{h}_{4}^{\ast }}{2%
\overline{h}_{4}}-\overline{h}_{3}^{\ast \ast }=-2\overline{h}_{3}\overline{h%
}_{4}\overline{\yen };  \label{eq2b} \\
\mathbf{R}_{3k}(\tau ) &=&\frac{\overline{h}_{3}}{2\overline{h}_{3}}%
\overline{n}_{k}^{\ast \ast }+(\frac{3}{2}\overline{h}_{3}^{\ast }-\frac{%
\overline{h}_{3}}{\overline{h}_{4}}\overline{h}_{4}^{\ast })\frac{\overline{n%
}_{k}^{\ast }}{2\overline{h}_{4}}=0;  \label{eq3b} \\
2\overline{h}_{3}\mathbf{R}_{4k}(\tau ) &=&-\overline{w}_{k}[\frac{\left(
\overline{h}_{3}^{\ast }\right) ^{2}}{2\overline{h}_{3}}+\frac{\overline{h}%
_{3}^{\ast }\overline{h}_{4}^{\ast }}{2\overline{h}_{4}}-\overline{h}%
_{3}^{\ast \ast }]+\frac{\overline{h}_{3}^{\ast }}{2}(\frac{\partial _{k}%
\overline{h}_{3}}{\overline{h}_{3}}+\frac{\partial _{k}\overline{h}_{4}}{%
\overline{h}_{4}})-\partial _{k}\overline{h}_{3}^{\ast }=0.  \label{eq4b}
\end{eqnarray}%
} This system of nonlinear PDEs can be transformed, respectively, into a
system of equation for stationary configurations if {\small
\begin{eqnarray}
\ _{h}\overline{\Upsilon }(\tau ,x^{i}) &\rightarrow &\ _{h}\Upsilon (\tau
,x^{i}),\newline
\overline{\Upsilon }(\tau ,x^{i},t)\rightarrow \Upsilon (\tau
,x^{i},y^{3}=\varphi ),  \label{station} \\
\overline{h}_{3}(\tau ,x^{i},t) &\rightarrow &h_{4}(\tau ,x^{i},\varphi ),%
\overline{h}_{4}(\tau ,x^{i},t)\rightarrow h_{3}(\tau ,x^{i},\varphi ),%
\overline{h}_{3}^{\ast }(\tau ,x^{i},t)\rightarrow h_{4}^{\diamond }(\tau
,x^{i},\varphi ),  \notag \\
\overline{h}_{4}^{\ast }(\tau ,x^{i},t) &\rightarrow &h_{3}^{\diamond }(\tau
,x^{i},\varphi ),\newline
\overline{w}_{k}(\tau ,x^{i},t)\rightarrow n_{k}(\tau ,x^{i},\varphi ),%
\overline{n}_{k}(\tau ,x^{i},t)\rightarrow w_{k}(\tau ,x^{i},\varphi )%
\mbox{
etc.}  \notag
\end{eqnarray}%
} Such a duality exists for Lorentz manifolds with a Killing symmetry when $%
y^{3}$ is a space like coordinate and $y^{4}=t$ is a time like coordinated.
This duality simplifies various applications of the AFDM when we can
redefine the procedure considered in the previous subsection for stationary
nonholonomic configurations to certain time dependent ones. It allows use to
prove decoupling properties and generate cosmological like solutions if
there are known certain stationary configurations, or inversely to find
certain stationary metrics as analogs of corresponding cosmological ones.%
\footnote{%
\ \label{lccondb}The LC-conditions (\ref{zerot}) for stationary
configurations transform into equations with coefficients depending on $t,$
\newline
$\partial _{t}\overline{w}_{i}=(\partial _{i}-\overline{w}_{i}\partial
_{t})\ln \sqrt{|\overline{h}_{4}|},(\partial _{i}-\overline{w}_{i}\partial
_{t})\ln \sqrt{|\overline{h}_{3}|}=0,\partial _{k}\overline{w}_{i}=\partial
_{i}\overline{w}_{k},\partial _{t}\overline{n}_{i}=0,\partial _{i}\overline{n%
}_{k}=\partial _{k}\overline{n}_{i}$. Such nonlinear first order PDEs
containing $\partial _{t}$ can be solved in explicit form for certain
classes of additional nonholonomic constraints on cosmological d--metrics
and N-coefficients, see (\ref{cosmf}).}

We can rewrite the nonlinear PDE (\ref{eq1b})--(\ref{eq4b}) in an explicit
decoupled form if we introduce the coefficients $\overline{\alpha }%
_{i}=(\partial _{t}\overline{h}_{3})\ (\partial _{i}\overline{\varpi }),\
\overline{\beta }=(\partial _{t}\overline{h}_{3})\ (\partial _{t}\overline{%
\varpi }),\ \overline{\gamma }=\partial _{t}\left( \ln |\overline{h}%
_{3}|^{3/2}/|\overline{h}_{4}|\right) ,$ where $\overline{\varpi }={\ln
|\partial _{t}}\overline{{h}}{_{3}/\sqrt{|\overline{h}_{3}\overline{h}_{4}|}|%
}.$ For $\partial _{t}h_{a}\neq 0$ and $\partial _{t}\varpi \neq 0,$ we
obtain such equations
\begin{equation}
\psi ^{\bullet \bullet }+\psi ^{\prime \prime }=2\ _{h}\overline{\yen };{%
\overline{\varpi }}^{\ast }\ \overline{h}_{3}^{\ast }=2\overline{h}_{3}%
\overline{h}_{4}\overline{\yen };\ \overline{n}_{i}^{\ast \ast }+\overline{%
\gamma }\overline{n}_{i}^{\ast }=0;\overline{\beta }\overline{w}_{i}-%
\overline{\alpha }_{i}=0.  \label{cosmsimpl}
\end{equation}%
We can integrate such equations "step by step" for any generating function $%
\overline{\Psi }(x^{i},t):=e^{\overline{\varpi }}$ and sources $\ _{h}%
\overline{\Upsilon }(x^{i})$ and $\overline{\Upsilon }(x^{k},t),$ see next
subsection. \vskip5pt

\textbf{Nonlinear symmetries for generating functions and sources with
effective cosmological constant:} The system (\ref{cosmsimpl}) with
respective coefficients relates four functions ($\overline{h}_{3},\overline{h%
}_{4},\overline{\Upsilon },\overline{\Psi })$ when a very important
nonlinear symmetry for locally anisotropic cosmological solutions and
respective generating functions, $(\overline{\Psi }(\tau ),\overline{%
\Upsilon }(\tau ))\iff (\overline{\Phi }(\tau ),\Lambda (\tau ))$ can be
found,
\begin{equation}
\overline{\Lambda }(\ \overline{\Psi }^{2})^{\ast }=|\overline{\yen }|(%
\overline{\Phi }^{2})^{\ast },\mbox{
or  }\overline{\Lambda }\overline{\ \Psi }^{2}=\overline{\Phi }^{2}|%
\overline{\yen }|-\int dt\ \overline{\Phi }^{2}|\overline{\yen }|^{\ast }.
\label{nsym1b}
\end{equation}%
This nonlinear symmetry allows us to introduce a new generating function $%
\overline{\Phi }({x}^{i},t)$ and an (effective) cosmological constant $%
\overline{\Lambda }(\tau )\neq 0$, which can be applied both for generating
exact off-diagonal solutions in explicit forms and elaborating on locally
anisotropic cosmological scenarios with cosmological constants, for
instance, considered in entropic gravity.

\subsection{Integrability of entropic quasiperiodic geometric flow equations}

We generate and study geometric properties of two classes of generic
off-diagonal solutions with elasticity and quasiperiodic structures of the
system of nonlinear PDEs (\ref{entropfloweq}). The first one is for
stationary configurations and the second one is considered for locally
anisotropic cosmological models.

\subsubsection{Off--diagonal cosmological solutions with elastic
quasiperiodic structures}

\label{ssintcosm}Applying the AFDM, we can construct cosmological solutions
(which, in general, are locally anisotropic) of the entropic flow modified
Einstein equations (\ref{entropfloweq}) with N-adapted sources \newline
$\ _{h}\overline{\yen }(\tau )=~\ _{h}\overline{\yen }(\tau ,x^{k})$ and $%
\overline{\yen }(\tau )=\overline{\yen }(\tau ,x^{k},t),$ see
parameterizations for cosmological configurations in \ (\ref{dsourcparam}).
Integrating "step by step" the system of the \ nonlinear PDEs (\ref{eq1b})--(%
\ref{eq4b}) decoupled in the form (\ref{cosmsimpl}), we obtain such
d--metric coefficients for (\ref{dm}), {\small
\begin{eqnarray}
\ g_{i}(\tau ) &=&e^{\ \psi (\tau ,x^{k})}%
\mbox{ as a solution of 2-d
Poisson eqs. }\psi ^{\bullet \bullet }+\psi ^{\prime \prime }=2~\ _{h}%
\overline{\yen }(\tau );  \notag \\
g_{3}(\tau ) &=&\overline{h}_{3}(\tau ,{x}^{i},t)=h_{3}^{[0]}(\tau
,x^{k})-\int dt\frac{(\overline{\Psi }^{2})^{\ast }}{4\overline{\yen }}%
=h_{3}^{[0]}(\tau ,x^{k})-\overline{\Phi }^{2}/4\overline{\Lambda }(\tau );
\notag \\
g_{4}(\tau ) &=&\overline{h}_{4}(\tau ,{x}^{i},t)=-\frac{(\overline{\Psi }%
^{2})^{\ast }}{4\overline{\yen }^{2}\overline{h}_{3}}=-\frac{(\overline{\Psi
}^{2})^{\ast }}{4\overline{\yen }^{2}(h_{3}^{[0]}(\tau ,x^{k})-\int dt(%
\overline{\Psi }^{2})^{\ast }/4\overline{\yen })}  \label{offdcosm} \\
&=&-\frac{[(\overline{\Phi }^{2})^{\diamond }]^{2}}{4\overline{h}_{3}|%
\overline{\Lambda }(\tau )\int dt\overline{\yen }[\overline{\Phi }%
^{2}]^{\diamond }|}=-\frac{[(\overline{\Phi }^{2})^{\ast }]^{2}}{%
4[h_{3}^{[0]}(x^{k})-\overline{\Phi }^{2}/4\overline{\Lambda }(\tau )]|\int
dt\ \overline{\yen }[\overline{\Phi }^{2}]^{\ast }|}.  \notag
\end{eqnarray}%
The N--connection coefficients are computed,
\begin{eqnarray}
N_{k}^{3}(\tau ) &=&\overline{n}_{k}(\tau ,{x}^{i},t)=\ _{1}n_{k}(\tau
,x^{i})+\ _{2}n_{k}(\tau ,x^{i})\int dt\frac{(\overline{\Psi }^{\ast })^{2}}{%
\overline{\yen }^{2}|h_{3}^{[0]}(\tau ,x^{i})-\int dt(\overline{\Psi }%
^{2})^{\ast }/4\overline{\yen }|^{5/2}}  \notag \\
&=&\ _{1}n_{k}(\tau ,x^{i})+\ _{2}n_{k}(\tau ,x^{i})\int dt\frac{(\overline{%
\Phi }^{\ast })^{2}}{4|\overline{\Lambda }(\tau )\int dt\overline{\yen }[%
\overline{\Phi }^{2}]^{\ast }||\overline{h}_{3}|^{5/2}};  \label{nconcosm} \\
N_{i}^{4}(\tau ) &=&\overline{w}_{i}(\tau ,{x}^{i},t)=\frac{\partial _{i}\
\overline{\Psi }}{\ \overline{\Psi }^{\ast }}=\frac{\partial _{i}\ \overline{%
\Psi }^{2}}{(\overline{\Psi }^{2})^{\ast }}\ =\frac{\partial _{i}[\int dt\
\overline{\yen }(\ \overline{\Phi }^{2})^{\ast }]}{\overline{\yen }(\
\overline{\Phi }^{2})^{\ast }},\ \   \notag
\end{eqnarray}%
} In these formulas, $h_{3}^{[0]}(\tau ,x^{k}),$ $\ _{1}n_{k}(\tau ,x^{i}),$
and $\ _{2}n_{k}(\tau ,x^{i})$ are integration functions encoding various
possible sets of (non) commutative parameters and integration constants
running on $\tau $ for geometric evolution flows. We can chose different
generating data $(\overline{\Psi }(\tau ,{x}^{i},t),\overline{\Upsilon }%
(\tau ,{x}^{i},t))$ or $(\overline{\Phi }(\tau ,{x}^{i},t),\overline{\Lambda}%
(\tau ))$ which are related by nonlinear differential / integral transforms (%
\ref{nsym1b}), and respective integration functions. Such values should be
chosen in explicit form following certain topology/ symmetry / asymptotic
conditions for some classes of exact / parametric cosmological solutions.
The coefficients (\ref{offdcosm}) and (\ref{nconcosm}) define generic
off-diagonal cosmological solutions if the corresponding anholonomy
coefficients are not trivial. Such locally cosmological solutions are with
nontrivial nonholonomically induced d-torsion and N-adapted coefficients
which can be computed in explicit form. In order to generate as particular
cases some well-known cosmological FLRW, or Bianchi, type metrics, we have
to consider data of type $(\overline{\Psi }(\tau ,t),\overline{\Upsilon }%
(\tau ,t)),$ or $(\overline{\Phi }(\tau ,t),\overline{\Lambda }(\tau )),$
with integration functions which allow frame/ coordinate transforms to
respective (off-) diagonal configurations $g_{\alpha \beta }(\tau ,t).$

Let us analyze certain important nonholonomic evolution properties of above
locally anisotropic cosmological solutions using the formulas for effective
sources (\ref{effsourc}) with cosmological parameterizations (\ref%
{dsourcparam}). In N-adapted form, we obtain a system of equations with
first order evolution derivatives $\partial _{\tau }$ when the v-part of
vierbeinds depend on a time like coordinate $y^{4}=t,$ $\mathbf{e}_{\ \mu
}^{\mu ^{\prime }}(\tau )=[\mathbf{e}_{\ 1}^{1^{\prime }}(\tau ,x^{k}),%
\mathbf{e}_{\ 2}^{2^{\prime }}(\tau ,x^{k}),$ \newline
$\mathbf{e}_{\ 3}^{3^{\prime}}(\tau ,x^{k},t),\mathbf{e}_{\ 4}^{4^{\prime
}}(\tau ,x^{k},t)] $; there are considered coordinates $(x^{k},t),$ when the
dependence on $y^{3}$ can be omitted because of Killing symmetry on $%
\partial _{3}.$ We can consider frame transforms for generating effective
sources, $\ ^{eff}\yen _{i}(\tau ) =[\mathbf{e}_{\ i}^{i^{\prime }}(\tau
)]^{2})[~\ ^{tot}\mathbf{\Upsilon }_{i^{\prime }1^{\prime }}(\tau )+\frac{1}{%
2} ~\partial _{\tau }\mathbf{g}_{1^{\prime }}(\tau )]=\ _{h}\yen (\tau ,{x}%
^{k}),$
\begin{equation}
\ ^{eff}\yen _{a}(\tau ) =[\mathbf{e}_{\ a}^{a^{\prime }}(\tau )]^{2})[~\
^{tot}\mathbf{\Upsilon }_{a^{\prime }a^{\prime }}(\tau )+\frac{1}{2}%
~\partial _{\tau }\mathbf{g}_{a^{\prime }}(\tau )]=\yen (\tau ,{x}^{k},t).
\label{cosmsourcevol}
\end{equation}%
In these formulas, we can prescribe any values for the matter sources $\
^{tot}\mathbf{\Upsilon }_{\mu \nu }(\tau )$ in a cosmological or spacetime
QC model. Then, for simplicity, we can consider N-adapted diagonal
configurations and integrate on $\tau $ and determine a cosmological
evolution flow of $\mathbf{g}_{\alpha ^{\prime }}(\tau ,{x}^{k},t)$ modelled
as a nonholonomic and nonlinear geometric diffusion process. All geometric
constructions are performed with respect to a new system of reference
determined by $\mathbf{e}_{\ \mu }^{\mu ^{\prime }}(\tau ,x^{k},t)$. We have
to prescribe some locally anisotropic generating values $[\mathbf{e}_{\ \mu
}^{\mu ^{\prime }}(\tau ,x^{k},t),\ ^{tot}\mathbf{\Upsilon }_{\mu \nu} (\tau
,x^{k},t)]$ which are compatible with certain observational data, for
instance, in modern cosmology and dark matter and dark energy physics.

\subsubsection{Qudratic line elements for off-diagonal cosmological
configurations with elastic flows}

\ Any coefficient $\overline{h}_{3}(\tau )=\overline{h}_{3}(\tau
,x^{k},t)=h_{3}^{[0]}(\tau ,x^{k})-\overline{\Phi }^{2}/4\overline{\Lambda }%
(\tau ),$ $\overline{h}_{3}^{\ast }\neq 0,$ can be considered also as a
generating function, for instance, for entropic quasiperiodic
configurations. Using formulas (\ref{offdcosm}), we find $\overline{\Phi }%
^{2}=-4\overline{\Lambda }(\tau )\overline{h}_{3}(\tau ,{r,\theta },t)$
transforming (\ref{nsym1b}) in $(\overline{\Psi }^{2})^{\ast }=\int dt\
\overline{\yen }\overline{h}_{3}^{\ast }.$ Introducing such values into the
formulas for $\overline{h}_{a}$ and $\overline{\yen }$ in (\ref{offdcosm})
and (\ref{nconcosm}), we construct locally anisotropic cosmological
solutions parameterized by \ d--metrics (\ref{dm}) with N-adapted
coefficients (\ref{cosmf}), {\small
\begin{eqnarray}
&&ds^{2}=e^{\ \psi (\tau ,x^{k})}[(dx^{1})^{2}+(dx^{2})^{2}]+
\label{gensolcosm} \\
&&\left\{
\begin{array}{cc}
\begin{array}{c}
\overline{h}_{3}[dy^{3}+(\ _{1}n_{k}+4\ _{2}n_{k}\int dt\frac{(\overline{h}%
_{3}^{\ast }{})^{2}}{|\int dy^{4}\ \overline{\yen }\overline{h}_{3}^{\ast
}|\ (\overline{h}_{3})^{5/2}})dx^{k}] \\
-\frac{(\overline{h}_{3}^{\ast }{})^{2}}{|\int dt\ \ \overline{\yen }%
\overline{h}_{3}^{\ast }|\ \overline{h}_{3}}[dt+\frac{\partial _{i}(\int dt\
\overline{\yen }\ \overline{h}_{3}^{\ast }{})}{\ \ \ \overline{\yen }%
\overline{h}_{3}^{\ast }\ {}}dx^{i}], \\
\mbox{ or }%
\end{array}
&
\begin{array}{c}
\mbox{gener.  funct.}\overline{h}_{3}, \\
\mbox{ source }\ \overline{\yen },\mbox{ or }\overline{\Lambda }(\tau );%
\end{array}
\\
\begin{array}{c}
(h_{3}^{[0]}-\int dt\frac{(\overline{\Psi }^{2})^{\ast }}{4\ \overline{\yen }%
})[dy^{3}+(_{1}n_{k}+\ _{2}n_{k}\int dt\frac{(\overline{\Psi }^{\ast })^{2}}{%
4\ \overline{\yen }^{2}|h_{3}^{[0]}-\int dy^{4}\frac{(\overline{\Psi }%
^{2})^{\ast }}{4\ \overline{\yen }}|^{5/2}})dx^{k}] \\
-\frac{(\overline{\Psi }^{2})^{\ast }}{4\ \overline{\yen }%
^{2}(h_{3}^{[0]}-\int dt\frac{(\overline{\Psi }^{2})^{\ast }}{4\ \overline{%
\yen }})}[dt+\frac{\partial _{i}\ \overline{\Psi }}{\ \overline{\Psi }^{\ast
}}dx^{i}], \\
\mbox{ or }%
\end{array}
&
\begin{array}{c}
\mbox{gener.  funct.}\overline{\Psi }, \\
\mbox{source }\ \overline{\yen };%
\end{array}
\\
\begin{array}{c}
(h_{3}^{[0]}-\frac{\overline{\Phi }^{2}}{4\overline{\Lambda }}%
)[dy^{3}+(_{1}n_{k}+\ _{2}n_{k}\int dt\frac{[(\overline{\Phi }^{2})^{\ast
}]^{2}}{|\ 4\overline{\Lambda }\int dy^{4}\ \overline{\yen }(\overline{\Phi }%
^{2})^{\ast }|}|h_{4}^{[0]}(x^{k})-\frac{\overline{\Phi }^{2}}{4\overline{%
\Lambda }}|^{-5/2})dx^{k}] \\
-\frac{[(\overline{\Phi }^{2})^{\ast }]^{2}}{|\ 4\overline{\Lambda }\int
dy^{4}\ \overline{\yen }(\overline{\Phi }^{2})^{\ast }|(h_{3}^{[0]}-\frac{%
\overline{\Phi }^{2}}{4\overline{\Lambda }})}[dt+\frac{\partial _{i}[\int
dt\ \ \overline{\yen }(\overline{\Phi }^{2})^{\ast }]}{\ \ \overline{\yen }(%
\overline{\Phi }^{2})^{\ast }}dx^{i}],%
\end{array}
&
\begin{array}{c}
\mbox{gener.  funct.}\overline{\Phi } \\
\overline{\Lambda }(\tau )\mbox{ for }\ \overline{\yen }.%
\end{array}%
\end{array}%
\right.  \notag
\end{eqnarray}%
} Such solutions posses a Killing symmetry on $\partial _{3}$ and can be
re-written in terms of $\eta $--polarization function functions for target
locally anistropic cosmological metrics $\ \widehat{\mathbf{g}}\mathbf{=}%
[g_{\alpha }=\eta _{\alpha }\mathring{g}_{\alpha },\ \eta _{i}^{a}\mathring{N%
}_{i}^{a}]$ encoding primary cosmological data $[\mathring{g}_{\alpha },%
\mathring{N}_{i}^{a}].$

\subsubsection{Off-diagonal Levi-Civita entropic and quasiperiodic
cosmological configurations}

\ We can extract and model entropic flow evolution of cosmological
spacetimes in GR. To satisfy the zero torsion conditions (\ref{zerot}), see
equations in footnote \ref{lccondb}, let us consider a special class of
generating functions and sources when, for instance, $\overline{\Psi }(\tau
)=\overline{\check{\Psi}}(\tau ,x^{i},t),$ when $(\partial _{i}\overline{%
\check{\Psi}})^{\ast }=\partial _{i}(\overline{\check{\Psi}}^{\ast })$ and $%
\overline{\yen }(\tau ,x^{i},t)=\overline{\yen }[\overline{\check{\Psi}}]=%
\overline{\check{\yen }}(\tau ),$ or $\overline{\yen }=const.$ For such
classes of entropic quasiperiodic generating functions and sources, the
nonlinear symmetries (\ref{nsym1b}) are written $\overline{\Lambda }(\tau )\
\overline{\check{\Psi}}^{2}=\overline{\check{\Phi}}^{2}|\overline{\check{%
\yen }}|-\int dt\ \overline{\check{\Phi}}^{2}|\overline{\check{\yen }}%
|^{\ast },\overline{\check{\Phi}}^{2}=-4\overline{\Lambda }(\tau )\overline{%
\check{h}_{3}}(\tau ,{r,\theta },t),$ $\overline{\check{\Psi}}^{2}=\int dt\
\overline{\check{\yen }}(\tau ,{r,\theta },t)\overline{\check{h}}_{3}^{\ast
}(\tau ,{r,\theta },t).$ Using these formulas, we conclude that the
coefficient $\overline{h}_{4}(\tau )=\overline{\check{h}}_{4}(\tau ,x^{i},t)$
can be considered also as generating function for entropic cosmological
solutions. For such LC--configurations, there are some parametric on $\tau $
functions $\overline{\check{A}}(\tau ,x^{i},t)$ and $n(\tau ,x^{i})$ when
the N--connection coefficients are computed
\begin{equation*}
\overline{n}_{k}(\tau )=\overline{\check{n}}_{k}(\tau )=\partial _{k}%
\overline{n}(\tau ,x^{i})\mbox{ and }\overline{w}_{i}(\tau )=\partial _{i}%
\overline{\check{A}}(\tau )=\frac{\partial _{i}(\int dt\ \overline{\check{%
\yen }}\ \overline{\check{h}}_{3}^{\ast }{}{}])}{\ \ \overline{\check{\yen }}%
\ \overline{\check{h}}_{3}^{\ast }{}}=\frac{\partial _{i}\overline{\check{%
\Psi}}}{\overline{\check{\Psi}}^{\ast }}=\frac{\partial _{i}[\int dt\ \
\overline{\check{\yen }}(\overline{\check{\Phi}}^{2})^{\ast }]}{\ \
\overline{\check{\yen }}(\overline{\check{\Phi}}^{2})^{\ast }}.
\end{equation*}

Summarizing above formulas, we construct new classes of locally anisotropic
cosmological solutions as ub GR defined as subclasses of solutions (\ref%
{gensolcosm}) with zero torsion but with entropic quasiperiodic geometric
flow evolution, {\small
\begin{eqnarray}
&&ds^{2}=e^{\ \psi (\tau ,x^{k})}[(dx^{1})^{2}+(dx^{2})^{2}]+
\label{lcsolcosm} \\
&&\left\{
\begin{array}{cc}
\begin{array}{c}
\overline{\check{h}}_{3}\left[ dy^{3}+(\partial _{k}\overline{n})dx^{k}%
\right] -\frac{(\ \overline{\check{h}}_{3}^{\ast }{}{}{})^{2}}{|\int dt\ \ \
\overline{\check{\yen }}\ \overline{\check{h}}_{3}^{\ast }{}|\ \ \overline{%
\check{h}}_{3}}[dt+(\partial _{i}\overline{\check{A}})dx^{i}], \\
\mbox{ or }%
\end{array}
&
\begin{array}{c}
\mbox{gener.  funct.}\overline{\check{h}}_{3}, \\
\mbox{ source }\ \overline{\check{\yen }},\mbox{ or }\overline{\Lambda };%
\end{array}
\\
\begin{array}{c}
(h_{3}^{[0]}-\int dt\frac{(\overline{\check{\Psi}}^{2})^{\ast }}{4\overline{%
\check{\yen }}})[dy^{3}+(\partial _{k}\overline{n})dx^{k}]-\frac{(\overline{%
\check{\Psi}}^{2})^{\ast }}{4\overline{\check{\yen }}^{2}(h_{3}^{[0]}-\int dt%
\frac{(\overline{\check{\Psi}}^{2})^{\ast }}{4\overline{\check{\yen }}})}%
[dt+(\partial _{i}\overline{\check{A}})dx^{i}], \\
\mbox{ or }%
\end{array}
&
\begin{array}{c}
\mbox{gener.  funct.}\overline{\check{\Psi}}, \\
\mbox{source }\ \overline{\check{\yen }};%
\end{array}
\\
(h_{3}^{[0]}-\frac{\overline{\check{\Phi}}^{2}}{4\overline{\Lambda }}%
)[dy^{3}+(\partial _{k}\overline{n})dx^{k}]-\frac{[(\overline{\check{\Phi}}%
^{2})^{\ast }]^{2}}{|\ 4\overline{\Lambda }\int dt\overline{\check{\yen }}(%
\overline{\check{\Phi}}^{2})^{\ast }|(h_{3}^{[0]}-\frac{\overline{\check{\Phi%
}}^{2}}{4\overline{\Lambda }})}[dt+(\partial _{i}\overline{\check{A}}%
)dx^{i}], &
\begin{array}{c}
\mbox{gener.  funct.}\ \overline{\check{\Phi}} \\
\mbox{effective }\overline{\Lambda }\mbox{ for }\ \overline{\check{\yen }}.%
\end{array}%
\end{array}%
\right.  \notag
\end{eqnarray}%
}

Such cosmological metrics are generic off-diagonal and define new classes of
solutions if the anholonomy coefficients are not zero for $N_{k}^{3}(\tau
=\partial _{k}\overline{n}$ and $N_{i}^{4}(\tau =\partial _{i}\overline{%
\check{A}}.$ They encode entropic quasiperiodic structures. We can analyze
certain nonholonomic cosmological configurations determined, for instance,
by data $(\overline{\check{\yen }},\overline{\check{\Psi}},h_{3}^{[0]},%
\overline{\check{n}}_{k}),$ when $\partial _{k}\overline{n}\rightarrow 0$
and $\overline{w}_{i}=\partial _{i}\overline{\check{A}}\rightarrow 0.$ Zero
values can be fixed also by certain additional nonholonomic constraints.
Choosing data $(\overline{\check{\yen }}(\tau ,t),\overline{\check{\Psi}}%
(\tau ,t),h_{3}^{[0]}=const,\overline{\check{n}}_{k}=const),$ we can
generate (off-) diagonal entropic metrics of Bianchi, or FLRW, types and
generalizations to other type configurations $g_{\alpha \beta }(\tau ,t)$ in
GR modified under geometric flow evolution.

\section{Entropic quasiperiodic flows and cosmological solutions}

\label{s5}The goal of this section is to consider physical implications of
models with entropic and quasiperiodic flow evolution of locally anisotropic
and inhomogeous cosmological spacetimes.

\subsection{The AFDM for entropic flow cosmological solutions}

We outline the key steps on the AFDM for generating cosmological solutions
with geometric flows and Killing symmetry on $\partial _{3}.$ Considering a
nonholonomic deformation procedure for a generating function $g_{3}(\tau )=%
\overline{h}_{3}(\tau ,x^{i},y^{3})$ (\ref{offdcosm}), cosmological
constants $\overline{\Lambda }(\tau )$ and sources $\ _{h}\overline{\yen }%
(\tau )=~\ _{h}\overline{\yen }(\tau ,x^{k})$ and $\overline{\yen }(\tau )=%
\overline{\yen }(\tau ,x^{k},t),$ see parameterizations (\ref{dsourcparam})
and nonlinear symmetries (\ref{nsym1b}), we construct exact solutions of the
system of nonlinear PDEs for emergent cosmology (\ref{cosmsimpl}).

Typical cosmological solutions of this class are parameterised {\small
\begin{eqnarray*}
ds^{2} &=&e^{\ \psi (\tau ,x^{k})}[(dx^{1})^{2}+(dx^{2})^{2}]+\overline{h}%
_{3}(\tau )[dy^{3}+(\ _{1}n_{k}+4\ _{2}n_{k}\int dt\frac{(\overline{h}%
_{3}^{\ast }{}(\tau ))^{2}}{|\int dt\ \overline{\yen }(\tau )\overline{h}%
_{3}^{\ast }|\ (\overline{h}_{3}(\tau ))^{5/2}})dx^{k}] \\
&&-\frac{[\overline{h}_{3}^{\ast }{}(\tau )]^{2}}{|\int dt\ \overline{\yen }%
(\tau )(h_{4}{}^{\diamond }(\tau ))|\ h_{4}}[dt+\frac{\partial _{i}(\int dt\
\overline{\yen }\ (\tau )\overline{h}_{3}^{\ast }{}(\tau ))}{\overline{\yen }%
(\tau )\ \overline{h}_{3}^{\ast }{}(\tau )}dx^{i}].
\end{eqnarray*}%
} Such quadratic line elements are time dual to the stationary ones which
can be obtained via transforms (\ref{station}).

\subsection{Nonlinear PDEs for entropic quasiperiodic cosmology}

We analyse two possibilities to transform the entropic flow modified
Einstein equations (\ref{entropfloweq}) into systems of nonlinear PDEs (\ref%
{eq1b})--(\ref{eq4b}) with generic off-diagonal or diagonal solutions
depending in explicit form on a evolution parameter, a time like variable
and two space like coordinates. In the first case, there are considered
entropic quasiperiodic sources determined by some additive or general
nonlinear functionals for effective matter fields. In the second case,
respective nonlinear functionals determining quasiperiodic solutions for
entropic configurations are prescribed for generating functions subjected to
nonlinear symmetries (\ref{nsym1b}). We also note that is is possible to
construct certain classes of locally anisotropic and inhomogeneous
cosmological solutions using nonlinear / additive functionals both for
generating functions and (effective) sources.

\subsubsection{Cosmological solutions for entropic quasiperiodic sources}

\paragraph{Cosmological configurations generated by additive entropic
functionals for sources:}

For this class of cosmological solutions, we consider an additive functional
for an entropic quasiperiodic source of type $\overline{\yen }(\tau
,x^{i},t) $ (\ref{dsourcparam}),%
\begin{equation}
\ ^{as}\overline{\yen }(\tau )=\ ^{as}\overline{\yen }(\tau ,{x}^{i},t)=\
^{fl}\overline{\yen }(\tau ,{x}^{i},t)+\ ^{m}\overline{\yen }(\tau ,{x}%
^{i},t)+\ \ ^{F}\overline{\yen }(\tau ,{x}^{i},t)+\ _{0}^{int}\overline{\yen
}(\tau ,{x}^{i},t)+\ _{0}^{\chi }\overline{\yen }(\tau ,{x}^{i},t).
\label{adsourccosm}
\end{equation}%
There is also an associated additive cosmological constant $\ ^{as}\overline{%
\Lambda }(\tau )$ (\ref{lcsolcosm}) related to different types of generating
functions via nonlinear symmetries (\ref{nsym1b}) when the equation (\ref%
{eq2b}) transforms into ${\overline{\varpi }}^{\ast }\ \overline{h}%
_{3}^{\ast }=2\overline{h}_{3}\overline{h}_{4}\ ^{as}\overline{\yen }(\tau )$
and can be integrated on time like variable $y^{4}=t.$ The systems of
nonlinear PDEs (\ref{cosmsimpl}) can be integrated following the AFDM
explained in details in \cite{v05,v09,v09a,gheorghiu14,bubuianu17,v16,
bubuianu18,vacaru18tc} (see sections on exact and parametric cosmological
solutions in MGTs) and \cite%
{vacaru06,vacaru07,vacaru09,vacaru10,vacaru13,ruchin13,gheorghiu16,rajpoot17,bubuianu18a}%
, for solutions with nonholonomic cosmological Ricci flows, and citations
therein. Such generic off-diagonal cosmological solutions are parameterized
in the form {\small
\begin{eqnarray}
&&ds^{2}=e^{\ \psi (\tau ,x^{k})}[(dx^{1})^{2}+(dx^{2})^{2}]+\overline{h}%
_{3}(\tau )[dy^{3}+(\ _{1}n_{k}(\tau ,{x}^{i})+4\ _{2}n_{k}(\tau ,{x}^{i})
\label{cosmasdm} \\
&&\int dt\frac{[\overline{h}_{3}{}^{\ast }(\tau )]^{2}}{|\int dt\ \ ^{as}%
\overline{\yen }(\tau )\overline{h}_{3}{}^{\ast }(\tau )|[\overline{h}%
_{3}(\tau )]^{5/2}})dx^{k}]-\frac{[\overline{h}_{3}{}^{\ast }(\tau )]^{2}}{%
|\int dt\ ^{as}\overline{\yen }(\tau )\overline{h}_{3}{}^{\ast }(\tau )|\
\overline{h}_{3}(\tau )}[dt+\frac{\partial _{i}(\int dt\ \ ^{as}\overline{%
\yen }(\tau )\ \overline{h}_{3}{}^{\ast }(\tau ))}{\ \ ^{as}\overline{\yen }%
(\tau )\ \overline{h}_{3}{}^{\ast }(\tau )}dx^{i}].  \notag
\end{eqnarray}%
} In such quadratic linear elements, we have to fix a sign of the
coefficient $\overline{h}_{3}(\tau ,x^{k},t)$ which describes relativistic
flow evolution with a generating function with Killing symmetry on $\partial
_{3}$ determined by sources $(\ _{h}\overline{\yen }(\tau ),\ ^{as}\overline{%
\yen }(\tau )).$ Such entropic and quasiperiodic flow solutions are of type (%
\ref{gensolcosm}) and can be re-written equivalently with coefficients
stated as functionals of $\ ^{as}\overline{\Phi }(\tau ,{x}^{i},t)$ and $\
^{as}\overline{\Psi }(\tau ,{x}^{i},t).$

We can extract from off-diagonal d-metrics (\ref{cosmasdm}) certain
cosmological LC-configurations determined by entropic quasiperiodic sources
by imposing additional zero torsion constraints. Such anholonomy conditions
restrict the respective classes of generating functions $(\overline{\check{h}%
}_{3}(\tau ,x^{i},t),\overline{\check{\Psi}}(\tau ,x^{i},t)$ and/or $%
\overline{\check{\Phi}}(\tau ,x^{i},t))$ for $\overline{n}(\tau , x^i),
\overline{\check{A}}(\tau ,{x}^{i},t))$ and sources $\ ^{as}\overline{\check{%
\yen }}(\tau )$ (\ref{adsourccosm}) and $\ ^{as}\overline{\Lambda }(\tau )$ (%
\ref{lcsolcosm}), {\small
\begin{equation}
ds^{2} =e^{\ \psi (\tau)}[(dx^{1})^{2}+(dx^{2})^{2}]+\overline{\check{h}}%
_{3}(\tau )\left[ dy^{3}+(\partial _{k}\overline{n}(\tau))dx^{k}\right] -%
\frac{[\overline{\check{h}}_{3}{}^{\ast }(\tau )]^{2}}{|\int dt\ ^{as}%
\overline{\check{\yen }}(\tau )\overline{\check{h}}_{3}{}^{\ast }(\tau )|\
\overline{\check{h}}_{3}(\tau )}[dt+(\partial _{i}\overline{\check{A}}%
(\tau))dx^{i}].  \label{cosmadmlc}
\end{equation}
}

The d-metrics (\ref{cosmasdm}) and/or (\ref{cosmadmlc}) define off-diagonal
cosmological solutions generated by entropic quasiperiodic additive sources $%
\ ^{as}\overline{\yen }(\tau )$ and/or $\ ^{as}\overline{\check{\yen }}(\tau
).$ The terms (\ref{adsourccosm}) encode and model respectively
contributions of standard and/or dark matter fields and effective entropic
evolution sources. Such values can be can be prescribed in certain forms
being compatible to observational data of cosmological (and
geometric/entropic) evolution for dark matter distributions with possible
quasiperiodic, aperiodic, pattern forming, solitonic nonlinear wave
interactions.

\paragraph{Cosmological solutions for nonlinear entropic quasiperiodic
functionals for sources:}

Such classes of exact cosmological solutions can be generated by nonlinear
quasiperiodic functionals for effective sources, $\ ^{qp}\overline{\yen }%
(\tau )=\quad ^{qp}\overline{\yen }(\tau ,x^{i},t)=\ ^{qp}\overline{\yen }[\
^{fl}\overline{\yen },\ ^{m}\overline{\yen },\ ^{F}\overline{\yen },\
_{0}^{int}\overline{\yen }+\ _{0}^{\chi }\overline{\yen }],$ see functional
dependencies in (\ref{adsourccosm}), subjected to nonlinear symmetries (\ref%
{nsym1b}). Applying the AFDM, we construct entropic flow cosmological
solutions of with nonlinear sources, {\small
\begin{eqnarray}
&&ds^{2}=e^{\ \psi (\tau )}[(dx^{1})^{2}+(dx^{2})^{2}]+\overline{h}_{3}(\tau
)[dy^{3}+(\ _{1}n_{k}(\tau )+4\ _{2}n_{k}(\tau )\int dt\frac{[\overline{h}%
_{3}{}^{\ast }(\tau )]^{2}}{|\int dt\ \ ^{qp}\overline{\yen }(\tau )%
\overline{h}_{3}{}^{\ast }(\tau )|[\overline{h}_{3}(\tau )]^{5/2}})dx^{k}]
\notag \\
&&-\frac{[\overline{h}_{3}{}^{\ast }(\tau )]^{2}}{|\int dt\ ^{qp}\overline{%
\yen }(\tau )\ \overline{h}_{3}{}^{\ast }(\tau )|\ \overline{h}_{3}(\tau )}%
[dt+\frac{\partial _{i}(\int dt\ \ ^{qp}\overline{\yen }(\tau )\ \overline{h}%
_{3}{}^{\ast }(\tau )])}{\ \ ^{qp}\overline{\yen }(\tau )\ \overline{h}%
_{3}{}^{\ast }(\tau )}dx^{i}].  \label{cosmnfdm}
\end{eqnarray}%
}

For LC-configurations, we obtain {\small
\begin{equation}
ds^{2} = e^{\ \psi (\tau )}[(dx^{1})^{2}+(dx^{2})^{2}]+\overline{\check{h}}%
_{3}(\tau) [dy^{3}+(\partial _{k}\overline{n}(\tau ))dx^{k}] -\frac{[%
\overline{\check{h}}_{3}{}^{\ast }(\tau )]^{2}}{|\int dt\ {\ ^{qp}}\overline{%
\check{\yen }}(\tau )\overline{h}_{3}{}^{\ast }(\tau )|\ \overline{\check{h}}%
_{3}(\tau )}[dt+(\partial _{i}\overline{\check{A}}(\tau))dx^{i}].
\label{cosmnfdmlcn}
\end{equation}%
} For additive functionals for cosmological entropic and quasiperiodic
sources, the formulas (\ref{cosmnfdm}) and (\ref{cosmnfdmlcn}) transforms
respectively into quadratic linear elements (\ref{cosmasdm}) and (\ref%
{cosmadmlc}). Fixing a value $\tau _{0},$ we obtain cosmological solutions
for Ricci solitons and MGTs.

\subsubsection{Cosmological configurations with nonstationary entropic
generating functions}

In this section, the sources for (effective) matter fields and geometric
flows are defined by arbitrary functions $\overline{\yen }_{\ \nu }^{\mu
}(\tau )=[~\ _{h}\overline{\yen }(\tau ,{x}^{k}),\overline{\yen }(\tau ,{x}%
^{k},t)].$ The quasiperiodic structure will be stated for some additive or
general nonlinear fuctionals for the generating functions.

\paragraph{Cosmological metrics with additive entropic generating functions:}

Such functionals are
\begin{equation}
\ ^{a}\overline{\Phi }(\tau )=\ ^{a}\overline{\Phi }(\tau ,{x}^{i},t)=\ ^{fl}%
\overline{\Phi }(\tau ,{x}^{i},t)+\ ^{m}\overline{\Phi }(\tau ,{x}^{i},t)+\
\ ^{F}\overline{\Phi }(\tau ,{x}^{i},t)+\ _{0}^{int}\overline{\Phi }(\tau ,{x%
}^{i},t)+\ _{0}^{\chi }\overline{\Phi }(\tau ,{x}^{i},t).
\label{aditcosmfunct}
\end{equation}%
The values $\ _{0}^{int}\overline{\Phi }(\tau )=\ _{0}^{int}\overline{\Phi }%
(\tau ,{x}^{i},t)=$ $\ ^{int}\Phi \lbrack \tau ,\varsigma ,\overline{b}]$
and $\ _{0}^{\chi }\overline{\Phi }(\tau )=$ $\ _{0}^{\chi }\overline{\Phi }%
(\tau ,{x}^{i},t)=\ ^{\chi }\overline{\Phi }[\tau ,\varsigma ,\overline{b}]$
are functionals on certain quasiperiodic (space time like QC or other type
aperiodic, solitonic structures) given by functions $\varsigma $ and/or $%
\overline{b}$ subjected to conditions of type (\ref{1tcqc}) and/or (\ref%
{evoleq}). The terms in the sum (\ref{aditcosmfunct}) correspond to the
Lagrange densities (\ref{lagrs}) and energy-momentum tensors (\ref{totsourc}%
) nonholonomically parameterised by sources (\ref{dsourcparam}). Changing
systems of references and coordinates, we can compute sums of functionals
for sources of type (\ref{adsourccosm}) using nonlinear symmetries (\ref%
{nsym1b}) considering an associated additive cosmological constant $\ ^{as}%
\overline{\Lambda }(\tau )$ (\ref{lcsolcosm}). For $\ ^{a}\overline{\Phi }%
(\tau )$ (\ref{aditcosmfunct}), the equation (\ref{eq2b}) transforms into a
functional equation $\overline{\varpi }^{\ast }(\tau )[\ ^{a}\overline{\Phi }%
(\tau ),\overline{\Lambda }(\tau )]\ \overline{h}_{3}^{\ast }(\tau )[\ ^{a}%
\overline{\Phi }(\tau ),\overline{\Lambda }(\tau )]=2\overline{h}_{3}(\tau
)[\ ^{a}\overline{\Phi }(\tau ),\overline{\Lambda }(\tau )]\overline{h}%
_{4}(\tau )[\ ^{a}\overline{\Phi }(\tau ),\overline{\Lambda }(\tau )]%
\overline{\yen }(\tau )$, and related analogs of (\ref{eq3b}) and (\ref{eq4b}%
) in the decoupled system of nonlinear PDEs (\ref{cosmsimpl}). Such
equations and their solutions can be written equivalently in different forms
with additive functionals of type $\ ^{a}\overline{h}_{3,4}$ and/or $\ ^{a}%
\overline{\Psi }$ and respective nonlinear functionals for the coefficients
in respective equations. Applying the AFDM as in the subsection \ref%
{ssintcosm}, we generate a class of parametric solutions of (\ref{cosmsimpl}%
) parameterized in a form similarly to (\ref{gensolcosm}),{\small
\begin{eqnarray}
&&ds^{2}=e^{\ \psi (\tau
,x^{k})}[(dx^{1})^{2}+(dx^{2})^{2}]+[h_{3}^{[0]}(\tau ,{x}^{i})-\frac{\ ^{a}%
\overline{\Phi }^{2}(\tau )}{4\overline{\Lambda }(\tau )}]  \notag \\
&&\left[ dy^{3}+\left( \ _{1}n_{k}(\tau ,{x}^{i})+\ _{2}n_{k}(\tau ,{x}%
^{i})\int dt\frac{[(\ ^{a}\overline{\Phi }^{2}){}^{\ast }(\tau )]^{2}}{4|%
\overline{\Lambda }(\tau )\int dt\ \overline{\yen }(\tau )(\ ^{a}\overline{%
\Phi }^{2}(\tau ))^{\ast }|(h_{3}^{[0]}(\tau ,{x}^{i})-\frac{\ ^{a}\overline{%
\Phi }^{2}(\tau )}{4\overline{\Lambda }(\tau )})^{5/2}}\right) dx^{k}\right]
\notag \\
&&-\frac{[(\ ^{a}\overline{\Phi }^{2}){}^{\ast }(\tau )]^{2}}{4|\overline{%
\Lambda }(\tau )\int dt\ \overline{\yen }(\tau )(\ ^{a}\overline{\Phi }%
^{2})^{\ast }(\tau )|\ [h_{3}^{[0]}(\tau ,{x}^{i})-\frac{\ ^{a}\overline{%
\Phi }^{2}(\tau )}{4\overline{\Lambda }(\tau )}]}[dt+\frac{\partial _{i}%
\left[ \int dt\ \ \overline{\yen }\ (\tau )(\ ^{a}\overline{\Phi }%
^{2}){}^{\ast }(\tau )\right] }{\ \overline{\yen }(\tau )\ (\ ^{a}\overline{%
\Phi }^{2}){}^{\ast }(\tau )}dx^{i}],  \label{cosmdmaf}
\end{eqnarray}%
} for integration functions $h_{3}^{[0]}(\tau ,{x}^{i}),\ _{1}n_{k}(\tau ,{x}%
^{i})$ and$\ _{2}n_{k}(\tau ,{x}^{i}).$

LC-configurations can be extracted from (\ref{cosmdmaf}) imposing additional
integrability conditions on coefficients resulting in zero nonholonomic
torsion. Such solutions can be considered both for relativistic entropic
Ricci solitons and in GR, being parameterized as in (\ref{lcsolcosm}),%
{\small
\begin{eqnarray}
ds^{2} &=&e^{\ \psi (\tau ,x^{k})}[(dx^{1})^{2}+(dx^{2})^{2}]
+[h_{3}^{[0]}(\tau ,{x}^{i})-\frac{\overline{\check{\Phi}}^{2}(\tau )}{4%
\overline{\Lambda }(\tau )}][dy^{3}+(\partial _{k}\overline{n}(\tau ,{x}%
^{i}))dx^{k}]  \label{cosmdmaflc} \\
&&-\frac{[(\ ^{a}\overline{\check{\Phi}}^{2}){}^{\ast }(\tau )]^{2}}{4|%
\overline{\Lambda }(\tau )\int dt\overline{\check{\yen }}(\tau )(\ ^{a}%
\overline{\check{\Phi}}^{2}){}^{\ast }(\tau )|\ (h_{3}^{[0]}(\tau ,{x}^{i})-%
\frac{\overline{\check{\Phi}}^{2}(\tau )}{4\overline{\Lambda }(\tau )})}%
[dt+(\partial _{i}\overline{\check{A}}(\tau ,{x}^{k},t))dx^{i}].  \notag
\end{eqnarray}
}

We can consider small parametric decompositions and frame/coordinate
transforms (in a more general context, we can elaborate a formalism of $\eta
$- and $\varepsilon $-polarization functions as for BH solutions but with
time like dependence for cosmological configurations ) in order to relate
new classes of solutons (\ref{cosmdmaf}) \ and/or (\ref{cosmdmaflc}) to some
well known (off-) diagonal cosmological metrics.

\paragraph{Cosmological configurations with nonlinear enropic functionals
for generating functions:}

Instead of additive generating functionals $\ ^{a}\overline{\Phi }(\tau )$ (%
\ref{aditcosmfunct}), we can work with nonlinear quasiperiodic generating
functionals $\ ^{qp}\overline{\Phi }(\tau )=\ \ ^{qp}\overline{\Phi }(\tau ,{%
x}^{i},t)=\ ^{qp}\overline{\Phi }[\ ^{fl}\overline{\Phi },\ ^{m}\overline{%
\Phi },\ ^{F}\overline{\Phi },\ _{0}^{int}\overline{\Phi },\ _{0}^{\chi }%
\overline{\Phi }]$ characterized by nonlinear symmetries of type (\ref%
{nsym1b}). The equation (\ref{eq2b}) transforms into a nonlinear functional
equation, \newline
$\overline{\varpi }^{\ast }(\tau )[\ ^{qp}\overline{\Phi }(\tau ),\overline{%
\Lambda }(\tau )]\ \overline{h}_{3}^{\ast }(\tau )[\ ^{qp}\overline{\Phi }%
(\tau ),\overline{\Lambda }(\tau )]=2\overline{h}_{3}(\tau )[\ ^{qp}%
\overline{\Phi }(\tau ),\overline{\Lambda }(\tau )]\overline{h}_{4}(\tau )[\
^{qp}\overline{\Phi }(\tau ),\overline{\Lambda }(\tau )]\ \overline{\yen }%
(\tau )$, which can be solved together with other equations with decoupling (%
\ref{cosmsimpl}). We obtain such solutions: {\small
\begin{eqnarray}
&&ds^{2}=e^{\ \psi (\tau ,x^{k})}[(dx^{1})^{2}+(dx^{2})^{2}]
+[h_{3}^{[0]}(\tau ,{x}^{i})-\ \frac{^{qp}\overline{\Phi }^{2}(\tau )}{4%
\overline{\Lambda }(\tau )}]  \label{cosmdmnf} \\
&& \lbrack dy^{3}+(\ _{1}n_{k}(\tau ,{x}^{i})+\ _{2}n_{k}(\tau ,{x}^{i})\int
dt\frac{[(\ ^{qp}\overline{\Phi }^{2}){}^{\ast }(\tau )]^{2}}{4|\overline{%
\Lambda }(\tau )\int dt\ \overline{\yen }(\tau )(\ ^{qp}\overline{\Phi }%
^{2})^{\ast }(\tau )|(h_{3}^{[0]}(\tau ,{x}^{i})-\ \frac{^{qp}\overline{\Phi
}^{2}(\tau )}{4\overline{\Lambda }(\tau )})^{5/2}})dx^{k}]  \notag
\end{eqnarray}
\begin{equation}
-\frac{\lbrack (\ ^{qp}\overline{\Phi }^{2}){}^{\ast }(\tau )]^{2}}{4|%
\overline{\Lambda }(\tau )\int dt\ \ \overline{\yen }(\tau )(\ ^{qp}%
\overline{\Phi }^{2})^{\ast }(\tau )|\ (h_{3}^{[0]}(\tau ,{x}^{i})-\frac{\
^{qp}\overline{\Phi }^{2}(\tau )}{4\overline{\Lambda }(\tau )})}[dt+\frac{%
\partial _{i}(\int dt\ \ \ \overline{\yen }(\tau )\ (\ ^{qp}\overline{\Phi }%
^{2}){}^{\ast }\ \overline{\yen }(\tau ))}{\ \overline{\yen }(\tau )\ (\
^{qp}\overline{\Phi }^{2}){}^{\ast }\ \overline{\yen }(\tau )}dx^{i}],
\notag
\end{equation}%
} where $h_{3}^{[0]}(\tau ,{x}^{i}),\ _{1}n_{k}(\tau ,{x}^{i})$ and$\
_{2}n_{k}(\tau ,{x}^{i})$ are integration functions.

For zero torsion constraints, we extract LC-configurations, {\small
\begin{eqnarray}
ds^{2} &=&e^{\ \psi (x^{k})}[(dx^{1})^{2}+(dx^{2})^{2}]+(h_{3}^{[0]}(\tau ,{x%
}^{i})-\ \frac{^{qp}\overline{\check{\Phi}}^{2}(\tau )}{4\overline{\Lambda }%
(\tau )}\ )[dy^{3}+(\partial _{k}\overline{n}(\tau ,{x}^{i}))dx^{k}]
\label{cosmdmnflc} \\
&&-\frac{[(\ ^{qp}\overline{\check{\Phi}}^{2}){}^{\ast }(\tau )]^{2}}{4|%
\overline{\Lambda }(\tau )\int dt\ \overline{\check{\yen }}(\tau )(\ ^{qp}%
\overline{\check{\Phi}}^{2})^{\ast }(\tau )|\ (h_{3}^{[0]}(\tau ,{x}^{i})-\
\frac{^{qp}\overline{\check{\Phi}}^{2}(\tau )}{4\overline{\Lambda }(\tau )})}%
[dt+(\partial _{i}\overline{\check{A}}(\tau ,{x}^{i},t))dx^{i}].  \notag
\end{eqnarray}%
} The coefficients of d--metrics (\ref{cosmdmnf}) and (\ref{cosmdmnflc}) can
be chosen to be of necessary smooth class and involve certain entropic,
quasiperiodic, stochastic sources and fractional derivative processes. Such
nonholonomic deformation and generalized transforms can be constructing with
changing the topological spacetime structure and modeling certain dark
energy and dark matter effects as results of entropic quasiperiodic flows or
nonholonomic deformations of certain prime cosmological solutions.

\subsubsection{Emergent quasiperiodic cosmology from both generating
functionals \& sources}

We can generate entropic flow cosmological solutions using generalized
quasiperiodic nonlinear functionals both for generating functions, $\ ^{qp}%
\overline{\Phi }(\tau ),$ and nonlinear functionals for (effective) sources,
$\ \ ^{qp}\overline{\yen }(\tau ),$ see above formulas. Such data are
connected via nonlinear symmetries of type (\ref{nsym1b}), when $\ ^{qp}%
\overline{\Lambda }(\tau )\ ^{qp}\overline{\Psi }^{2}(\tau )=\ ^{qp}%
\overline{\Phi }^{2}(\tau )|\ ^{qp}\overline{\yen }(\tau )|-\int dt\ \ ^{qp}%
\overline{\Phi }^{2}(\tau )|\ ^{qp}\overline{\yen }(\tau )|^{\ast }.$
Similar nonlinear symmetries exist for additive functionals both for the
gravitational fields and (effective) sources, with (\ref{aditcosmfunct}) and
(\ref{adsourccosm}) and can be written in a particular form $\ ^{a}\overline{%
\Lambda \ }(\tau )\ ^{a}\overline{\Psi }^{2}(\tau )=\ ^{a}\overline{\Phi }%
^{2}|\ ^{a}\overline{\yen }(\tau )|-\int dt\ ^{a}\overline{\Phi }^{2}|\ ^{a}%
\overline{\yen }(\tau )|^{\ast }.$ Applying the AFDM\ summarized in Table 3
(using data $(\ ^{qp}\overline{\Psi }(\tau ),\ ^{qp}\overline{\yen }(\tau
)), $ and/or, equivalently, $(\ \ ^{qp}\overline{\Phi }(\tau ),\ ^{qp}%
\overline{\Lambda }(\tau ))),$ we construct multi-functional nonlinear
entropic quasiperiodic cosmological configurations,
\begin{eqnarray}
&&ds^{2} =e^{\ \psi (\tau
,x^{k})}[(dx^{1})^{2}+(dx^{2})^{2}]+(h_{3}^{[0]}(\tau ,{x}^{i})-\frac{^{qp}%
\overline{\Phi }^{2}(\tau )}{4\ ^{qp}\overline{\Lambda }(\tau )})
\label{cosmsdmnfg} \\
&&[dy^{3}+(\ _{1}n_{k}(\tau ,{x}^{i})+\ _{2}n_{k}(\tau ,{x}^{i})\int dt\frac{%
[(\ ^{qp}\overline{\Phi }^{2}){}^{\ast }(\tau )]^{2}}{4\ ^{qp}|\overline{%
\Lambda }(\tau )\int dt\ \ ^{qp}\overline{\yen }(\ ^{qp}\overline{\Phi }%
^{2})^{\ast }|(h_{3}^{[0]}(\tau ,{x}^{i})-\frac{^{qp}\overline{\Phi }%
^{2}(\tau )}{4\ ^{qp}\overline{\Lambda }(\tau )})^{5/2}})dx^{k}]  \notag \\
&&-\frac{[(\ ^{qp}\overline{\Phi }^{2}){}^{\ast }(\tau )]^{2}}{4|\ ^{qp}%
\overline{\Lambda }(\tau )\int dt\ \ ^{qp}\overline{\yen }(\tau )(\ ^{qp}%
\overline{\Phi }^{2})^{\ast }|\ (h_{3}^{[0]}(\tau ,{x}^{i})-\frac{^{qp}%
\overline{\Phi }^{2}(\tau )}{4\ ^{qp}\overline{\Lambda }(\tau )})}[dt+\frac{%
\partial _{i}(\int dt\ \ \ ^{qp}\overline{\yen }(\tau )\ (\ ^{qp}\overline{%
\Phi }^{2}){}^{\ast }(\tau )])}{\ \ ^{qp}\overline{\yen }(\tau )\ (\ ^{qp}%
\overline{\Phi }^{2}){}^{\ast }(\tau )}dx^{i}],  \notag
\end{eqnarray}%
where for integration functions $h_{3}^{[0]}(\tau ,{x}^{i}),\ _{1}n_{k}(\tau
,{x}^{i})$ and$\ _{2}n_{k}(\tau ,{x}^{i}).$

For LC-configurations, we obtain multi-functional nonlinear generalizations
of (\ref{cosmdmnf}) and (\ref{cosmdmnflc}) modelling locally anisotropic and
inhomogeneous solutions in entropic flow gravity and GR,%
\begin{eqnarray}
ds^{2} &=&e^{\ \psi (\tau
,x^{k})}[(dx^{1})^{2}+(dx^{2})^{2}]+(h_{3}^{[0]}(\tau ,{x}^{i})-\frac{\ ^{qp}%
\overline{\check{\Phi}}^{2}(\tau )}{4\ ^{qp}\overline{\Lambda }(\tau )}%
)[dy^{3}+(\partial _{k}\overline{n}(\tau ,{x}^{i},t))dx^{k}]
\label{cosmdmnflcg} \\
&&-\frac{[(\ ^{qp}\overline{\check{\Phi}}^{2}){}^{\ast }(\tau )]^{2}}{4|\
^{qp}\overline{\Lambda }(\tau )\int dt\ \ ^{qp}\ \overline{\check{\yen }}%
(\tau )(\ ^{qp}\overline{\check{\Phi}}^{2})^{\ast }(\tau )|\
(h_{3}^{[0]}(\tau ,{x}^{i})-\frac{\ ^{qp}\overline{\check{\Phi}}^{2}(\tau )}{%
4\ ^{qp}\overline{\Lambda }(\tau )})}[dt+(\partial _{i}\overline{\check{A}}%
(\tau ,{x}^{i},t))dx^{i}].  \notag
\end{eqnarray}

The classes of cosmological solutions (\ref{cosmsdmnfg}) and (\ref%
{cosmdmnflcg}) describe off-diagonal entropic non-stationary configurations
determined by multi-functional nonlinear quasiperiodic structures.

\subsection{Cosmological metrics evolving in entropic quasiperiodic media}

Generic off-diagonal entropic and quasiperiodic cosmological solutions are
constructed in terms of $\eta $--polarization functions following the AFDM
method for parametric $\tau $ and time like depending evolution. We consider
a primary cosmological d-metric $\mathbf{\mathring{g}}$ (\ref{primedm})
defined by data $[\mathring{g}_{i}(x^{k},t),\mathring{g}_{a}=\mathring{h}%
_{a}(x^{k},t);\mathring{N}_{k}^{3}=\mathring{n}_{k}(x^{i},t),\mathring{N}%
_{k}^{4}=\mathring{w}_{k}(x^{i},t)]$ which can be diagonalized for a FLRW
cosmological metric by frame/ coordinate transforms\footnote{%
in general, we can consider off-diagonal Bianchi anisotropic cosmological
metrics or any cosmological solution in GR or MGTs}. The cosmological
entropic quasiperiodic y target metrics $\overline{\mathbf{g}}(\tau )$ of
type (\ref{dme}) are generated by nonholonomic deformations
\begin{equation*}
\mathbf{\mathring{g}}\rightarrow \overline{\mathbf{g}}(\tau )\mathbf{=}[%
\overline{g}_{i}(\tau ,x^{k})=\overline{\eta }_{i}(\tau ,x^{k},t)\mathring{g}%
_{i},\overline{g}_{b}(\tau ,x^{k},t)=\overline{\eta }_{b}(\tau ,x^{k},t)%
\mathring{g}_{b},\overline{N}_{i}^{a}(\tau ,x^{k},t)=\ \overline{\eta }%
_{i}^{a}(\tau ,x^{k},t)\mathring{N}_{i}^{a}],
\end{equation*}%
when overlined symbols are used for distinguishing cosmological d-metrics
from stationary ones studied in previous sections. The quadratic line
elements corresponding to target locally anisotropic and inhomogeneous
cosmological metrics $\overline{\mathbf{g}}(\tau )=\overline{\mathbf{g}}%
(\tau ,x^{k},t)$ can be written in terms of gravitational polarization
functions, {\small
\begin{eqnarray*}
&&ds^{2}=\overline{\eta }_{1}(\tau ,x^{i},t)\mathring{g}%
_{1}(x^{i},t)[dx^{1}]^{2}+\overline{\eta }_{2}(\tau ,x^{i},t)\mathring{g}%
_{2}(x^{i},t)[dx^{1}]^{2} \\
&&+\overline{\eta }_{3}(\tau ,x^{i},t)\mathring{h}_{3}(x^{i},t)[dy^{3}+%
\overline{\eta }_{i}^{3}(\tau ,x^{i},t)\mathring{N}%
_{i}^{3}(x^{k},t)dx^{i}]^{2}+\overline{\eta }_{4}(\tau ,x^{i},t)\mathring{h}%
_{4}(x^{i},t)[dt+\overline{\eta }_{i}^{4}(\tau ,x^{k},t)\mathring{N}%
_{i}^{4}(x^{k},t)dx^{i}]^{2}.
\end{eqnarray*}%
} The $\eta $-coefficients will be constructed in some explicit forms
determined by entropic quasiperiodics generation functions and/or effective
sources as solutions of the system of nonlinear PDEs (\ref{cosmsimpl}).

\subsubsection{Cosmological evolutions generated by nonstationary entropic
sources}

We consider entropic sources of type $\ ^{qp}\overline{\yen }(\tau )=\quad
^{qp}\overline{\yen }(\tau ,x^{i},t)=\ ^{qp}\overline{\yen }[\ ^{fl}%
\overline{\yen },\ ^{m}\overline{\yen },\ ^{F}\overline{\yen },\ _{0}^{int}%
\overline{\yen }+\ _{0}^{\chi }\overline{\yen }]$ as in (\ref{cosmnfdm}) and
(\ref{cosmnfdmlcn}) and compute the $\eta $--polarization functions
following formulas {\small
\begin{eqnarray}
\overline{\eta }_{i}(\tau ) &=&e^{\ \psi (\tau ,x^{i})}/\mathring{g}_{i};%
\mbox{  generating function}\overline{\eta }_{3}(\tau )=\overline{\eta }%
_{3}(\tau ,x^{i},t);\overline{\eta }_{4}(\tau )=-\frac{4[(|\overline{\eta }%
_{3}(\tau )\mathring{h}_{3}|^{1/2})^{\ast }]^{2}}{\mathring{h}_{4}|\int dt\
\ ^{qp}\overline{\yen }(\tau )(\overline{\eta }_{3}(\tau )\mathring{h}%
_{3})^{\ast }|\ };  \label{cosmpfqp} \\
\overline{\eta }_{i}^{3}(\tau ) &=&\frac{_{1}n_{k}(\tau ,x^{i})}{\mathring{n}%
_{k}}+4\frac{\ _{2}n_{k}(\tau ,x^{i})}{\mathring{n}_{k}}\int dt\frac{\left(
[(\overline{\eta }_{3}(\tau )\mathring{h}_{3})^{-1/4}]^{\ast }\right) ^{2}}{%
|\int dt\ \ ^{qp}\overline{\yen }(\tau )(\overline{\eta }_{3}(\tau )%
\mathring{h}_{3})^{\ast }|\ };\ \overline{\eta }_{k}^{4}(\tau )=\frac{%
\partial _{i}\ \int dt\ \ ^{qp}\overline{\yen }(\tau )(\overline{\eta }%
_{3}(\tau )\mathring{h}_{3})^{\ast }}{\overline{\mathring{w}}_{i}\ \ ^{qp}%
\overline{\yen }(\tau )(\overline{\eta }_{3}(\tau )\mathring{h}_{3})^{\ast }}%
,  \notag
\end{eqnarray}%
} where $\ _{1}n_{k}(\tau ,x^{i})$ and $\ _{2}n_{k}(\tau ,x^{i})$ are
integration functions.

Using $\overline{\eta }_{3}(\tau ,x^{i},t)$ as a generating function, we can
compute other types of generating functions of the same target cosmological
d-metric subjected to nonlinear symmetries (\ref{nsym1b}),
\begin{equation*}
\overline{\Phi }^{2}(\tau )=4|\ ^{qp}\overline{\Lambda }(\tau
)[h_{3}^{[0]}(\tau ,x^{k})-\overline{\eta }_{3}(\tau ,{x}^{i},t)\mathring{h}%
_{3}({x}^{k},t)]|,\ (\overline{\Psi }^{2})^{\ast }(\tau )=-\int dt\ \ ^{qp}%
\overline{\yen }(\tau )\ [\overline{\eta }_{3}(\tau ,x^{i},t)\overline{%
\mathring{h}}_{3}(x^{i},t)]^{\ast }.
\end{equation*}

For integrable generating functionals and sources, when the constructions (%
\ref{lccondb}) are subjected to target off-diagonal cosmological metrics (%
\ref{lcsolcosm}) with zero torsion, we obtain
\begin{eqnarray}
\overline{\eta }_{i}(\tau ) &=&e^{\ \psi (\tau ,x^{i})}/\mathring{g}_{i};%
\overline{\eta }_{3}(\tau )=\overline{\check{\eta}}_{3}(\tau ,{x}^{i},t)%
\mbox{  as
a generating function};  \label{cosmpfqplc} \\
\overline{\eta }_{4}(\tau ) &=&-\frac{4[(|\ \overline{\check{\eta}}_{3}(\tau
)\mathring{h}_{3}|^{1/2})^{\ast }]^{2}}{\mathring{h}_{4}|\int dt\ ^{qp}\
\overline{\check{\yen }}(\tau )(\overline{\check{\eta}}_{3}\mathring{h}%
_{3})^{\ast }|\ };\overline{\eta }_{k}^{3}(\tau )=\frac{\partial _{k}%
\overline{n}(\tau ,x^{i})}{\mathring{n}_{k}};\overline{\eta }_{k}^{4}(\tau )=%
\frac{\partial _{k}\overline{\check{A}}(\tau ,x^{i},t)}{\mathring{w}_{k}}.
\notag
\end{eqnarray}%
In (\ref{cosmpfqp}) and (\ref{cosmpfqplc}), the nonlinear functionals for
the entropic quasiperiodic v-source and (effective) cosmological constant
can be changed into additive functionals $\ ^{qp}\overline{\yen }(\tau
)\rightarrow \ ^{as}\overline{\yen }(\tau )$ and $\ ^{qp}\overline{\Lambda }%
(\tau )\rightarrow \ ^{as}\overline{\Lambda }(\tau )$ which generates
another classes of cosmological solutions.

\subsubsection{Cosmology from nonstationary entropic generating functions}

We can construct locally anisotropic and inhomogeneous cosmological
solutions as nonholonomic deformations of some prime cosmological metrics
when the coefficients of the d-metrics are determined nonlinear generating
functionals $\ ^{qp}\overline{\Phi }(\tau )=\ \ ^{qp}\overline{\Phi }(\tau ,{%
x}^{i},t)=\ ^{qp}\overline{\Phi }[\ ^{fl}\overline{\Phi },\ ^{m}\overline{%
\Phi },\ ^{F}\overline{\Phi },\ _{0}^{int}\overline{\Phi },\ _{0}^{\chi }%
\overline{\Phi }]$ as for (\ref{cosmdmnf}). It is possible also to generate
similar cosmological metrics by additive functionals $\ ^{a}\overline{\Phi }%
(\tau )$ (\ref{aditcosmfunct}) for prescribed families of effective sources $%
\overline{\yen }(\tau )=\overline{\yen }(\tau ,x^{i},t)$ and cosmological
constants $\overline{\Lambda }(\tau ).$ Using formulas for nonlinear
symmetries (\ref{nsym1b}), we compute (recurrently) corresponding nonlinear
functionals, $\ ^{qp}\overline{\eta }_{3}(\tau ,{x}^{i},t),$ or additive
functionals,$\ ^{a}\overline{\eta }_{3}(\tau ,{x}^{i},t),$ and related
polarization functions,{\small
\begin{equation*}
\ ^{qp}\overline{\Phi }^{2}(\tau ) = 4|\ \overline{\Lambda }(\tau
)[h_{3}^{[0]}(\tau ,x^{k})-\ ^{qp}\overline{\eta }_{3}(\tau ,{x}^{i},t)%
\mathring{h}_{3}({x}^{k},t)]|,\ (\ ^{qp}\overline{\Psi }^{2})^{\ast }(\tau )
= -\int dt\ \ \overline{\yen }(\tau )\ [\ ^{qp}\overline{\eta }_{3}(\tau
,x^{i},t)\overline{\mathring{h}}_{3}(x^{i},t)]^{\ast }.
\end{equation*}
}

The coefficients of quadratic elements of type (\ref{gensolcosm}) are
recurrently computed,{\small
\begin{eqnarray}
\overline{\eta }_{i}(\tau ) &=&e^{\ \psi (\tau ,x^{i})}/\mathring{g}_{i};%
\overline{\eta }_{3}(\tau )=\ ^{qp}\overline{\eta }_{3}(\tau ,x^{i},t)%
\mbox{
as a generating function};  \label{cosmpfqp1} \\
\overline{\eta }_{4}(\tau ) &=&-\frac{4[(|\ ^{qp}\overline{\eta }_{3}(\tau )%
\mathring{h}_{3}|^{1/2})^{\ast }]^{2}}{\mathring{h}_{4}|\int dt\overline{%
\yen }(\tau )(\ ^{qp}\overline{\eta }_{3}(\tau )\mathring{h}_{3})^{\ast }|\ }%
;  \notag \\
\overline{\eta }_{i}^{3}(\tau ) &=&\frac{_{1}n_{k}(\tau ,x^{i})}{\mathring{n}%
_{k}}+4\frac{\ _{2}n_{k}(\tau ,x^{i})}{\mathring{n}_{k}}\int dt\frac{\left(
[(\ ^{qp}\overline{\eta }_{3}(\tau )\mathring{h}_{3})^{-1/4}]^{\ast }\right)
^{2}}{|\int dt\overline{\yen }(\tau )(\ ^{qp}\overline{\eta }_{3}(\tau )%
\mathring{h}_{3})^{\ast }|\ };\ \overline{\eta }_{k}^{4}(\tau ) = \frac{%
\partial _{i}\ \int dt\ \overline{\yen }(\tau )(\ ^{qp}\overline{\eta }%
_{3}(\tau )\mathring{h}_{3})^{\ast }}{\overline{\mathring{w}}_{i}\ \overline{%
\yen }(\tau )(\ ^{qp}\overline{\eta }_{3}(\tau )\mathring{h}_{3})^{\ast }},
\notag
\end{eqnarray}%
} where $\ _{1}n_{k}(\tau ,x^{i})$ and $\ _{2}n_{k}(\tau ,x^{i})$ are
integration functions.

Target off-diagonal cosmological metrics (\ref{lcsolcosm}) with zero torsion
extracted from (\ref{cosmpfqp1}) can be generated by polarization functions%
{\small
\begin{eqnarray*}
\overline{\eta }_{i}(\tau ) &=&e^{\ \psi (\tau ,x^{i})}/\mathring{g}_{i};%
\mbox{  generating
function }\overline{\eta }_{3}(\tau )=\ ^{qp}\overline{\check{\eta}}%
_{3}(\tau ,{x}^{i},t); \\
\overline{\eta }_{4}(\tau ) &=&-\frac{4[(|\ ^{qp}\ \overline{\check{\eta}}%
_{3}(\tau )\mathring{h}_{3}|^{1/2})^{\ast }]^{2}}{\mathring{h}_{4}|\int dt%
\overline{\check{\yen }}(\tau )(\ ^{qp}\overline{\check{\eta}}_{3}(\tau )%
\mathring{h}_{3})^{\ast }|\ };\ \overline{\eta }_{k}^{3}(\tau ) = (\partial
_{k}\overline{n}(\tau ,x^{i}))/\mathring{n}_{k};\overline{\eta }%
_{k}^{4}=\partial _{k}\overline{\check{A}}(\tau ,{x}^{i},t)/\mathring{w}_{k}.
\end{eqnarray*}
}

The cosmological solutions generated in this subsection describe entropic
flow nonholonomic deformations of prime cosmological configurations (for
instance, a FLRW, or Bianchi, type metric, and various modifications in
accelerating cosmology) self-consistently imbedded into a quasiperiodic
gravitational (dark energy) media.

\subsubsection{Cosmological configurations for entropic sources \&
generating functions}

We can construct more general classes of nonholonomic deformations of prime
cosmological metrics generated by entropic quasiperiodic flow nonlinear
quaisperiodic functionals both for the generating functions and (effective)
sources. For such locally anisotropic and inhomogeneous cosmological models
defined by nonlinear superpositions of cosmological solutions (\ref{cosmpfqp}%
) and (\ref{cosmpfqp1}) when the coefficients of (\ref{cosmnfdm}) are
computed,{\small
\begin{equation*}
\overline{\eta }_{i}(\tau ) = e^{\ \psi (\tau ,x^{i})}/\mathring{g}_{i}; %
\mbox{generating function }\ ^{qp}\overline{\eta }_{3}(\tau )=\ ^{qp}%
\overline{\eta }_{3}(\tau ,x^{i},t); \overline{\eta }_{4}(\tau ) = -\frac{%
4[(|\ ^{qp}\overline{\eta }_{3}(\tau )\mathring{h}_{3}|^{1/2})^{\ast }]^{2}}{%
\mathring{h}_{4}|\int dt\ \ ^{qp}\overline{\yen }(\tau )\ (\ ^{qp}\overline{%
\eta }_{3}(\tau )\mathring{h}_{3})^{\ast }|\ };
\end{equation*}
\begin{equation}
\overline{\eta }_{i}^{3}(\tau ) = \frac{_{1}n_{k}(\tau)}{\mathring{n}_{k}}+%
\frac{4\ _{2}n_{k}(\tau)}{\mathring{n}_{k}}\int dt\frac{\left( [(\ ^{qp}%
\overline{\eta }_{3}(\tau )\mathring{h}_{3})^{-1/4}]^{\ast }\right) ^{2}}{%
|\int dt\ \ ^{qp}\overline{\yen }(\tau )\ (\ ^{qp}\overline{\eta }_{3}(\tau )%
\mathring{h}_{3})^{\ast }|\ }; \overline{\eta }_{k}^{4}(\tau ) = \frac{%
\partial _{i}\ \int dt\ \ \ ^{qp}\overline{\yen }(\tau )\ (\ ^{qp}\overline{%
\eta }_{3}(\tau )\mathring{h}_{3})^{\ast }}{\overline{\mathring{w}}_{i}\ \
^{qp}\overline{\yen }(\tau )\ (\ ^{qp}\overline{\eta }_{3}(\tau )\mathring{h}%
_{3})^{\ast }},  \label{cosmpfqp12}
\end{equation}%
} where $\ _{1}n_{k}(\tau ,x^{i})$ and $\ _{2}n_{k}(\tau ,x^{i})$ are
integration functions. In formulas (\ref{cosmpfqp12}), there are considered
nonlinear generating functionals $\ ^{qp}\overline{\Phi }(\tau )=\ \ ^{qp}%
\overline{\Phi }(\tau ,{x}^{i},t)=\ ^{qp}\overline{\Phi }[\ ^{fl}\overline{%
\Phi },\ ^{m}\overline{\Phi },\ ^{F}\overline{\Phi },\ _{0}^{int}\overline{%
\Phi },\ _{0}^{\chi }\overline{\Phi }]$ characterized by nonlinear
symmetries (\ref{nsym1b}) for some prescribed families of nonlinear
functionals $\ ^{qp}\overline{\yen }(\tau )=\quad ^{qp}\overline{\yen }(\tau
,x^{i},t)=\ ^{qp}\overline{\yen }[\ ^{fl}\overline{\yen },\ ^{m}\overline{%
\yen },\ ^{F}\overline{\yen },\ _{0}^{int}\overline{\yen }+\ _{0}^{\chi }%
\overline{\yen }]$ $\ $and $\ ^{qp}\overline{\Lambda }(\tau ).$ Instead of
nonlinear superpositions $(\ ^{qp}\overline{\Phi }(\tau ),\ ^{qp}\overline{%
\yen }(\tau ),\ ^{qp}\overline{\Lambda }(\tau )),$ we can consider additive
data $(\ ^{a}\overline{\Phi }(\tau ),\ ^{a}\overline{\yen }(\tau ),\ ^{a}%
\overline{\Lambda }(\tau )).$ $\ $We can define also general nonlinear, $\
^{qp}\overline{\eta }_{3}(\tau ,x^{i},t),$ or additive functionals, $\ ^{a}%
\overline{\eta }_{3}(\tau ,x^{i},t),$ for other types polarization
functions, {\small
\begin{equation*}
\ ^{qp}\overline{\Phi }^{2}(\tau ) =4|\ \ ^{qp}\overline{\Lambda }(\tau
)[h_{3}^{[0]}(\tau)-\ ^{qp}\overline{\eta }_{3}(\tau ,{x}^{i},t)\mathring{h}%
_{3}({x}^{k},t)]|\ , (\ ^{qp}\overline{\Psi }^{2})^{\ast }(\tau ) = -\int
dt\ \ \ ^{qp}\overline{\yen }(\tau )\ [\ ^{qp}\overline{\eta }_{3}(\tau
,x^{i},t)\overline{\mathring{h}}_{3}(x^{i},t)]^{\ast }.
\end{equation*}
}

There are defined LC-configurations with zero torsion for target
off-diagonal cosmological metrics (\ref{lcsolcosm}) if there are considered
integrable polarization functions{\small
\begin{eqnarray*}
\overline{\eta }_{i}(\tau ) &=&e^{\ \psi (\tau ,x^{i})}/\mathring{g}_{i};%
\mbox{ generating function
}\overline{\check{\eta}}_{3}(\tau )=\ ^{qp}\overline{\check{\eta}}_{3}(\tau ,%
{x}^{i},t); \\
\overline{\eta }_{4}(\tau ) &=&-\frac{4[(|\ ^{qp}\overline{\check{\eta}}%
_{3}(\tau )\mathring{h}_{3}|^{1/2})^{\ast }]^{2}}{\mathring{h}_{4}|\int dt\
^{qp}\overline{\check{\yen }}(\tau )(\ ^{qp}\overline{\check{\eta}}_{3}(\tau
)\mathring{h}_{3})^{\ast }|\ };\ \overline{\eta }_{k}^{3}(\tau ) = (\partial
_{k}\overline{n}(\tau ,x^{i}))/\mathring{n}_{k};\overline{\eta }%
_{k}^{4}(\tau )=\partial _{k}\overline{\check{A}}(\tau ,{x}^{i},t)/\mathring{%
w}_{k}.
\end{eqnarray*}
}

Finally, it should be noted that there is duality on $y^{3}$ and $y^{4}$
coordinates and respective N-connection coefficients for the class of
cosmological solutions (\ref{cosmpfqp12}) and the stationary solutions which
can be generated via transforms (\ref{station}).

\section{Conclusions and Discussion}

\label{s6}

In this article we elaborate a geometric approach to E. Verlinde conjecture
\cite{verlinde10,verlinde16} that gravity can be considered as emergent
phenomena of a conventional spacetime elasticity determined by certain
entropic forces. We argue that rigorous thermodynamic formulations exist for
the models of "entropic spacetime and gravity" which can be derived from
generalizations of the Poincar\'{e}--Thurston conjecture extended to
relativistic geometric flow evolution theories. This is possible if
corresponding nonholonomic modifications of Perelman's (entropic type)
functionals are performed. For self-similar configurations, certain type
entropic flow evolution equations result into corresponding nonholonomic
Ricci soliton equations which are equivalent to the motion equations in
emergent modified gravity theories, MGTs, and (for certain conditions) in
general relativity, GR.

To study geometric flow evolution of Riemannian metrics G. Perelman
introduced two Lyapunov type functionals (F- and W-entropy) \cite{perelman1}
which was a very important step to the proof of the Poincar\'{e} conjecture
and elaborating geometric and statistical thermodynamics models for Ricci
flows and possible applications in modern physics. In a series of our works
\cite%
{vacaru07,vacaru09,vacaru10,vdiffusion,vacaru13,ruchin13,gheorghiu16,rajpoot17,bubuianu18a}%
, we considered various modifications of the F- and W-entropy functionals
for constructing (non) commutative and/or supersymmetric geometric evolution
models; investigating nonlinear (fractional and/or locally anisotropic)
stochastic and kinetic processes in curved spaces; and elaborating on
cosmological scenarios encoding memory and quantum interactions of
gravitational and (effective) matter fields. Here we note that the concepts
of complex manifolds, supermanifolds and noncommutative geometry, are
defined by geometric constructions with nonholonomic distributions on curved
spaces. In such a general context, for various entropic spacetime and
emergent gravity models, and in geometric flow evolution theories with
nonholonomic constraints, we have to develop an unified geometric formalism
for metric--affine spaces, generalized Finsler-Lagrange-Hamilton geometry,
almost K\"{a}hler and noncommutative geometries etc. Such constructions were
performed for Ho\v{r}ava-Lifshits, $f(R),\ R^{2}$, and other types MGTs (see
reviews \cite{nojiri17,capozziello,hossain15,basilakos13,elghozi15}) with
developments for models of thermodynamic / entropic / entanglement and
emergent gravity \cite%
{jacobson15,padmanabhan09,verlinde10,verlinde16,hossenfelder17,dai17a,dai17b,raamsdonk10, lashkari13,ryu06,ryu06a,lloyd12,faulkner14,swingle12,pastawski15,solodukhin11}%
; and for locally anisotropic kinetic and thermodynamic theories on curved
spaces (see \cite{vacaru00ap,vdiffusion,ruchin13,ruppeiner,quevedo,castro19}
and references therein) etc.

We have found that using certain classes of nonholonomic variables the
(relativistic) geometric flow evolution and Ricci soliton equations, and
related motion equations in entropic and other type MGTs, can be decoupled
and integrated in some very general forms. This allows us to construct
various classes of exact and parametric solutions with generic off-diagonal
metrics and generalized connections. The coefficients of new classes of
stationary and (in general, locally anisotropic and inhomogeneous)
cosmological solutions depend on all spacetime and associated phase space
(kinetic and/or thermodynamic) coordinates via generating and integration
functions and various types of commutative and noncommutative parameters and
integration and physical constants. Such geometric and analytic techniques
of constructing exact solutions in geometric flow evolution and MGTs have
been developed in the framework of the so-called anholonomic frame
deformation method, AFDM \cite{v05,v09,v09a}. For details, examples of exact
solutions, and various applications, we cite \cite%
{bubuianu18a,bubuianu18,gheorghiu14,bubuianu17,v16,vacaru06,ruchin13,
gheorghiu16,rajpoot17,bubuianu18a} and references therein.

It should be noted that there were not elaborated corresponding topological
methods and a well-defined analytic formalism for investigating geometric
evolution equations of metrics of pseudo-Euclidean signature. A number of
such conceptual and fundamental issues in nonlinear functional analysis and
the geometry of Lorentz manifolds have not addressed or solved by
mathematicians. The standard geometric flow paradigm was proposed as the
Hamilton-Perleman program for Riemannian metrics defining entropy type
functionals and deriving nonlinear evolution equations. To elaborate
realistic relativistic physical models we deal with Ricci tensors which in
the limit of weak gravitational/ elastic flows approximate to the d'Alambert
(wave) operator and not to the Laplace (diffusion) one used for Euclidean
signatures. So, the original approach to the topology and geometric flows of
Riemannian metrics has to be generalized in certain relativistic and
nonholonomic forms which are compatible with modern experimental particle
physics data and observational cosmology. The Poincar\'{e}--Thurston
conjecture can be formulated and proven for non-relativistic evolution of
any 3-d space like hypersurface. In this and partner \cite{partner1} works,
we advocate that using E. Verlinde conjecture on elastic emergent gravity,
we can elaborate on generalizations of nonholonomic Ricci flow theories as
certain models of relativistic flow evolution. Such models are determined by
extensions of Perelman's functionals for 3-d Riemannian metrics to certain
modified 4-d F- and W-entropy nonholonomic analogs which are extended on a
time like coordinate and/or a temperature like evolution parameter. For
certain approaches with rich gravitational vacuum structure, we work with
geometric relativistic kinetic/ hydrodynamic/ thermodynamic models (see
details in \cite{ruchin13}) or (for instance, in this work) with a $\tau $%
--parametric theory describing geometric entropic flows determined by
relativistic and elastically spacetime modified nonholonomic F- and
W-functionals.

As it was mentioned above, there is not yet formulated a rigorous
mathematical approach to the theory relativistic of geometric/ entropic
flows of metrics with Lorentz signature and generalized connections.
Nevertheless, we shown that such theories are characterized by certain
classes of generalized R. Hamilton equations with effective parametric
sources which may encode entropic and quasiperiodic structures (these are
necessary for explaining, for instance, the complex structure of dark matter
and energy in modern cosmology). Using the AFDM, we proved that such
geometric flow evolution equations and their Ricci soliton variants can be
decoupled and integrated in very general forms. In section \ref{s5}, we
constructed and analyzed possible physical implications of respective
entropic locally anisotropic cosmological solutions. A series of our former
results on astrophysical and cosmological models with quasiperiodic, pattern
forming, quasicrystal time like structures, see \cite%
{v16,vacaru18tc,bubuianu17,amaral17,aschheim18,bubuianu18} and references
therein, were used for developing in this work similar models for the
entropic geometric flow and gravity theories. Such exact and parametric
solutions provide also explicit examples of entropic gravity models
developed in a phenomenological manner in \cite%
{verlinde10,verlinde16,hossenfelder17}. So, we conclude that the E. Verlinde
conjecture on entropic character of gravity can be related to a relativistic
extension of the Poincar\'{e}--Thurston conjecture. Even such geometric
ideas have not been proven as explicit theorems in modern geometric
analysis, there are rigorous exact solutions of respective systems of
nonlinear PDES which support such entropic gravity and flow evolution ideas.

In sections \ref{s4} and \ref{s5}, we shown how to construct in explicit
form entropic and quaisiperiodic solutions for relativistic geometric flows,
nonholonomic Ricci solitons and generalized gravitational field equations.
Such a techniques was elaborated similarly to MGTs with quasiperiodic
structure (constructed and studied in this and our partner works \cite%
{partner1,bubuianu18,vacaru18tc}) and involves generic off-diagonal metric
and nonholonomically deformed non-Riemannian linear and nonlinear
connections. We emphasized, see also \cite{bubuianu18a}, that such
configurations are not characterized, in general, by certain entropy-area,
holographic or duality conditions. As a consequence, it is not possible to
elaborate on thermodynamic models of entropic MGTs and physical properties
of their exact or parametric solutions using only the concepts related to
the Bekenstein-Hawking entropy. We consider that there is an alternative and
more general way when stationary and cosmological solutions in geometric/
entropic flow evolution theories, MGTs and GR, can be defined and
characterized by nonholonomic deformations of Perelman's W-entropy. Such
constructions are similar to the well-known results on relativistic locally
anisotropic thermodynamics and kinetics \cite{ruchin13,vacaru00ap} and can
be generalized for emergent classical and quantum gravity theories.

Finally, we note that our geometric/ entropic flow approach to MGTs provides
new mathematical methods and applications in the theory of classical and
quantum informatics, for research of quantum systems with entanglement,
models of quantum and emergent gravity, and accelerating cosmology and dark
energy/ matter interactions etc. Our further research programs are related
to developments in such directions.

\vskip3pt

\textbf{Acknowledgments:} This research develops authors' former research
programs on geometric flows and applications in physics and information theory partially supported by a fellowship at IMAFF CSIC Madrid; a visit at Fields Institute, Toronto; a
project IDEI, PN-II-ID-PCE-2011-3-0256; fellowships at CERN and DAAD
programs for W. Heisenberg (M. Plank) Institute, Munich.

\end{document}